\documentclass[prx, superscriptaddress, nofootinbib]{revtex4-2}

\usepackage{graphicx}
\usepackage{amsmath}
\usepackage{amssymb}
\usepackage{hyperref}
\hypersetup{
 pdfnewwindow=true,
 colorlinks=true,
 linkcolor=blue,
 citecolor=blue,
 filecolor=blue,
 urlcolor=blue
}
\usepackage{color}
\usepackage[caption=false, justification=centerlast]{subfig}
\usepackage{bm}
\usepackage{bbm}
\usepackage[normalem]{ulem}

\DeclareMathOperator{\tr}{\mathrm{tr}}
\DeclareMathOperator{\erf}{\mathrm{erf}}

\newcommand{\rp}{\mathrm{p}}
\newcommand{\rs}{\mathrm{s}}

\newcommand{\bea}{\begin{eqnarray}}
\newcommand{\eea}{\end{eqnarray}}
\newcommand{\besa}{\begin{subequations}\begin{eqnarray}}
\newcommand{\eesa}{\end{eqnarray} \end{subequations}}
\newcommand{\beaa}{\begin{eqnarray}\begin{aligned}}
\newcommand{\eeaa}{\end{aligned}\end{eqnarray}}

\newcommand{\av}[1]{\langle #1 \rangle}
\newcommand{\braket}[1]{\langle #1 \rangle}
\newcommand{\bra}[1]{\langle #1 |}
\newcommand{\ket}[1]{| #1 \hspace{0.1 mm} \rangle}

\newcommand{\comments}[1]{}

\newcommand{\mr}[1]{\mathrm{ #1 }}
\newcommand{\mc}[1]{\mathcal{ #1 }}

\newcommand{\id}{\mathbbm{I}}
\renewcommand{\d}{\mathrm{d}}

\begin{document}

\title{Dynamical symmetrization of the state of identical particles}

\author{Armen E. Allahverdyan}
\affiliation{A. Alikhanyan National Science Laboratory (YerPhI), 0036 Yerevan, Armenia}

\author{Karen V. Hovhannisyan}
\affiliation{The Abdus Salam International Centre for Theoretical Physics (ICTP), 34151 Trieste, Italy}
\affiliation{A. Alikhanyan National Science Laboratory (YerPhI), 0036 Yerevan, Armenia}

\author{David Petrosyan}
\affiliation{A. Alikhanyan National Science Laboratory (YerPhI), 0036 Yerevan, Armenia}
\affiliation{Institute of Electronic Structure and Laser, FORTH, GR-70013 Heraklion, Crete, Greece}

\begin{abstract}
We propose a dynamical model for state symmetrization of two identical particles produced in spacelike-separated events by independent sources. We adopt the hypothesis that the pair of non-interacting particles can initially be described by a tensor product state since they are in principle distinguishable due to their spacelike separation. As the particles approach each other, a quantum jump takes place upon particle collision, which erases their distinguishability and projects the two-particle state onto an appropriately (anti)symmetrized state. The probability density of the collision times can be estimated quasi-classically using the Wigner functions of the particles' wavepackets, or derived from fully quantum mechanical considerations using an appropriately adapted time-of-arrival operator. Moreover, the state symmetrization can be formally regarded as a consequence of the spontaneous measurement of the collision time. We show that symmetric measurements performed on identical particles can in principle discriminate between the product and symmetrized states. Our model and its conclusions can be tested experimentally.
\end{abstract}

\maketitle

\section{Introduction}

The symmetrization postulate of quantum mechanics states that any joint state of identical particles (i.e., having all their non-dynamic features, such as mass, charge, spin, etc., the same) should be either symmetric (for bosons) or antisymmetric (for fermions) under permutations of the particles \cite{Girardeau1965, Flicker1967, Salzman1970, Peres, Landau}. Common sense though suggests that we could, in principle, attach labels and distinguish identical, non-interacting particles emanating from different, largely separated sources, at least until their coordinate probability densities start to overlap. Indeed, assume that two distant sources $\mathcal{S}_L$ and $\mathcal{S}_R$ produce nearly-simultaneously two identical particles 1 and 2 in states $\ket{L}$ and $\ket{R}$, respectively, in spacelike-separated events. Then, if a particle is detected close to a source, say $\mathcal{S}_L$, shortly after the creation, we can conclude with high degree of confidence that this is the same particle 1 that was emitted by $\mathcal{S}_L$. The hypothesis that independently generated, spatially separated identical particles can be considered distinguishable, with the joint wavefunction represented by a tensor product of the wavefunctions of each particle, has also been discussed in the past \cite{Mirman1973, Gelfer1975, Muynck1986}.
It was then argued that transition to the symmetrized state might occur once the spatial wavefunctions of the particles start to overlap, reducing their distinguishability \cite{Gelfer1975, Dieks1990}.  
It has recently been suggested that such a transition during bound state formation may be related to the excess energy transferred to an environment \cite{YungerHalpern2019}.

Let us recall two common arguments against the hypothesis that the initial state of largely separated particles can be represented by a tensor product of single-particle states. The first argument is field-theoretical \cite{Dieks1990}: particles correspond to excitations of a quantum field, and particle creation and annihilation correspond to a transition between the Fock states of the field effected by appropriate bosonic or fermionic field operators. This reasoning is, however, circular: the symmetrization postulate is not derived from second quantization; rather, the symmetrization postulate is used to derive the properties (commutation relations) of the second-quantized bosonic or fermionic field operators \cite{Green1953, Landau}\footnote{In this context, recall the message of the spin-statistics theorem from relativistic field theory \cite{Green1953, Ignatiev1987}: integer-spin (half-integer-spin) fields cannot be quantized via anticommutators (commutators), but can be quantized via commutators (anticommutators). Hence, integer-spin (half-integer-spin) fields can be represented as bosons (fermions). This statement however does not exclude that, e.g., half-integer spins can be quantized via some other algebraic (paraparticle) structure \cite{Green1953, Ignatiev1987}.}.

The second argument relies on the vast experimental evidence that identical particles within a small distance from each other, or those that have interacted with each other in the past, are in appropriately symmetrized states. Yet, for spatially separated, non-interacting particles, the results of localized in space measurements (i.e., coordinate, but not momentum) are the same for both product and symmetrized states \cite{Peres}. One may then argue that product states, even if they were possible, are redundant in the theory. These arguments, however, ignore the possibility of a transition between product and symmetrized states of non-interacting particles prepared far apart and approaching each other. Then, the difference between product and symmetrized states can be detected via symmetric measurements on the particles, even if their states do not (yet) overlap in space, as discussed below. 

Before proceeding, we note that measurements on identical particles are insensitive to their permutations, since the interaction of a measuring apparatus with identical particles is, by definition, symmetric under particle permutations \cite{Muynck1986, Dieks1990, Peres}. Hence, all allowed measurements on identical particles should be described by permutation-invariant operators \cite{Dieks1990, Peres}. Such measurements produce permutation-invariant results for any state, symmetrized or not. In fact, some textbooks do not account for this aspect and ``derive'' the state symmetrization from the symmetry of the (position) measurement statistics \cite{Landau, Messiah}. The constraint of permutation-symmetric observables for identical particles has important consequences for the definition of their reduced (marginal) states, as we discuss in Appendix~\ref{app:margo}.

\begin{figure}[t]
\centering
\includegraphics[width = 0.6 \columnwidth]{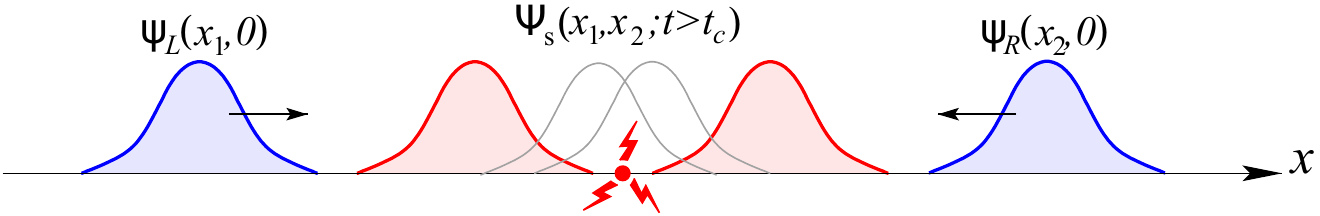}
\caption{Schematic representation of our 1D model. Two identical particles, $1$ and $2$, emanating at time $t = 0$ from different, spatially separated sources, have the coordinate wavefunctions $\psi_L(x_1, t) = \braket{x_1 | L(t)}$ and $\psi_R(x_2, t) = \braket{x_2 | R(t)}$ (blue) moving towards each other along the $x$ axis. The two-particle wavefunction $\Psi_{\mr{p}}(x_1,x_2,t) = \psi_L(x_1, t)\psi_R(x_2, t)$ factorizes. As their probability densities start to overlap (gray), the particles collide (red circle) at some time $t_c$, which results in the (anti)symmetrized two-particle wavefunction $\Psi_{\mr{s}}(x_1, x_2, t > t_c) \propto \psi_L(x_1, t) \psi_R(x_2, t) + \eta \psi_R(x_1, t) \psi_L(x_2, t)$ ($\eta = \pm 1$).}
\label{fig:2wps}
\end{figure}

In this paper, we adopt the initial product state hypothesis and show that permutation-symmetric measurements can in principle differentiate between the product and symmetrized states of the particles. We then propose a simple and intuitive one-dimensional (1D) dynamical model for symmetrization of the wavefunction of a pair of initially separated identical particles upon their collision, illustrated in Fig.~\ref{fig:2wps}. Inspired by the quantum jump approach to continuously measured quantum systems \cite{Dalibard1992, Dum1992, Gardiner1992, Plenio1998, Lambropoulos2007}, in our model, we apply the symmetrization operator to the two-particle state, which projects the initial product state onto a fully (anti)symmetric superposition. 
The physical intuition behind the application of the symmetrization operation is that, once the particles collide, they lose their individuality, which erases the possibility to distinguish their labels. Note that this happens operationally already for classical identical particles. 
We model the interaction between the particles similarly to how it is done for ideal gases: 
the particles are free \textit{except} when they are in an immediate neighborhood of each other.

Our symmetrization quantum jump occurs when the particles collide with each other, and we determine 
the probability density of the collision times from semiclassical and fully quantum-mechanical considerations. 
For semiclassical calculations of the collision time distribution, we assume Gaussian wavepackets for single particles and use the resulting non-negative Wigner functions \cite{Peres, Hudson1974}.
The semiclassical probability distribution of the collision times turns out to be nearly identical to that obtained with the fully quantum treatment of the collision time, which we determine by introducing an appropriate ``time of arrival'' operator \cite{Aharonov1961, Grot1996, Muga1998}. This operator is, however, not an ordinary self-adjoint operator corresponding to a standard projective measurement, but it can be described by a positive operator-valued measure (POVM) with the post-measurement state chosen arbitrarily. This permits us to formally regard the symmetrization quantum jump as a transition induced by a spontaneous measurement of the collision time. 

The paper is organized as follows. In Sec.~\ref{sec:mp} we review the symmetric measurements performed on a system of identical particles and show that such measurements can in fact discern the difference between the symmetrized and product states. In Sec.~\ref{sec:sy} we present our model for state symmetrization upon collision of the particles. In Sec.~\ref{sec:tc} we derive the probability density of collision times from the semiclassical and fully quantum perspectives. We summarize our model and the results in Sec.~\ref{sec:concl}. Some of the mathematical details are deferred to Appendices \ref{app:margo}, \ref{app:degeneracy}, \ref{app:time}, \ref{app:povm}, \ref{app:Gauss}, \ref{app:3D}.

\section{Measurement probabilities for identical particles}
\label{sec:mp}

Here we discuss which measurements are allowed for identical particles and how to employ them to discern the difference between product and symmetrized states. Clarifying these issues is important for understanding the physics of identical particles and avoiding some common misconceptions. As mentioned above, any Hamiltonian for a system of identical particles, including their interactions with measuring apparata, is symmetric under particle permutations. Hence, all allowed measurements on identical particles are described by permutation-symmetric operators \cite{Peres, Muynck1986, Dieks1990}, and this argument is independent on the state of the system (recall that in quantum mechanics measurement operators are defined independently from states). We emphasize that the hypothesized asymmetric product state $\ket{L} \otimes \ket{R}$ of identical particles is prepared as a result of particle generation, which is a process outside the formalism of non-relativistic quantum mechanics. Moreover, non-symmetric states cannot be prepared via projective measurements, since all such measurements are described by symmetric projectors.

Consider a single-particle operator $\hat{O}$ decomposed into a complete set of $N$ projectors
$\hat{P}_k$ as $\hat{O} = \sum_{k = 1}^N \alpha_k \hat{P}_k$, with 
\begin{equation} 
\hat{P}_k \hat{P}_l = \delta_{k l} \hat{P}_k, \quad \sum_{k = 1}^N \hat{P}_k = \id.
\end{equation} 
For a pair of particles, the permutation-symmetric measurement of operator 
$\hat{O} \otimes \hat{O} = \sum_{k \leq l} \alpha_k \alpha_l \hat{P}_{k l}$ then involves $N (N + 1)/2$ projectors
\beaa 
\label{symproj}
\hat{P}_{k l} \equiv \{ \hat{P}_k \otimes \hat{P}_l + \hat{P}_l \otimes \hat{P}_k\}_{l < k}^N , & 
\quad
\hat{P}_{k k} \equiv \{ \hat{P}_k \otimes \hat{P}_k\}_{k = 1}^N , 
\\
\sum_{k \leq l}^N \hat{P}_{k l} = \id \otimes \id. &
\eeaa
It follows that, for any density operator $\hat{\sigma}$ of the system, the measurement probabilities 
are permutation-symmetric:
\begin{equation}
\tr (\hat{\sigma} \hat{P}_{k l}) = \tr (\hat{\pi} \hat{\sigma} \hat{\pi}\, \hat{P}_{k l}),
\end{equation} 
where $\hat{\pi}$ is the particle permutation operator.

\subsection{Coordinate measurements}

Given a state $\ket{\Psi}$ of two identical particles living in a 1D space,
the joint probability density for the two coordinates reads
\begin{equation} 
D_{\Psi}(x_1, x_2) = \bra{\Psi} \hat{P}_{x_1 x_2} \ket{\Psi}, \label{lalo1}  
\end{equation}
with 
\begin{equation} \label{lalo2}
\hat{P}_{x_1 x_2} \equiv \ket{x_1} \bra{x_1} \otimes \ket{x_2} \bra{x_2} + \ket{x_2} \bra{x_2} \otimes \ket{x_1} \bra{x_1}, \quad \hat{P}_{x x} = \ket{x} \bra{x} \otimes \ket{x} \bra{x}
\end{equation}
being a continuous analogue of the projectors in Eq.~\eqref{symproj}. This representation stems from the decomposition $\hat{x} \otimes \hat{x} = \iint\limits_{x_1 \leq x_2} \d x_1 \d x_2 \, x_1 x_2 \hat{P}_{x_1 x_2}$ and the corresponding normalization condition $\iint\limits_{x_1 \leq x_2} \d x_1 \d x_2 \, \hat{P}_{x_1 x_2} = \id \otimes \id$.

Consider the two-particle product state
\begin{equation}
\label{eq:Psiproduct}
\ket{\Psi_{\mr{p}}} = \ket{L} \otimes \ket{R}
\end{equation}
with the first (second) particle produced by $\mathcal{S}_L$ ($\mathcal{S}_R$),
and the symmetric ($\eta = 1$) or antisymmetric ($\eta = -1$) state
\begin{equation}
\ket{\Psi_{\mr{s}}} = \frac{1}{\sqrt{2}}(\ket{L} \otimes \ket{R} + \eta \ket{R} \otimes \ket{L} ). 
\label{eq:Psisym}
\end{equation}
Using Eq.~\eqref{lalo1}, we obtain
\bea \label{c2}
D_{\Psi_\rp}(x_1, x_2) &=& |\psi_L(x_1)|^2 \, |\psi_R(x_2)|^2 + |\psi_L(x_2)|^2 \, |\psi_R(x_1)|^2,
\\ \nonumber
D_{\Psi_\rs}(x_1, x_2) &=& \bra{\Psi_\rs} \hat{P}_{x_1 x_2} \ket{\Psi_\rs} = \bra{\Psi_\rp} \hat{P}_{x_1 x_2} + \eta \hat{P}_{x_1 x_2} \hat{\pi} \ket{\Psi_\rp}
\\ \label{c5}
&=& D_{\Psi_\rp}(x_1, x_2) + \eta [\psi^*_L(x_1) \psi_L(x_2) \psi_R(x_1) \psi^*_R(x_2) + \mathrm{c.c.}],
\eea
where the coordinate wavefunctions are defined as
\begin{equation} \label{psidef}
\psi_{L, R}(x) \equiv \braket{x | L, R}.
\end{equation}
Thus, both $D_{\Psi_\rp}(x_1, x_2)$ and $D_{\Psi_\rs}(x_1, x_2)$ are invariant under permuting $x_1$ and $x_2$, even though $\ket{\Psi_\rp}$ is a non-symmetric product state. 
When the wavefunctions do not overlap in space, $\psi_L^* (x) \psi_R(x) \simeq 0 \, \forall \, x$, the second term on the right-hand side of Eq.~\eqref{c5} vanishes, and therefore $D_{\Psi_\rp}(x_1, x_2) \simeq D_{\Psi_\rs}(x_1, x_2)$, i.e., the non-symmetric product state $\ket{\Psi_\rp}$ of Eq.~\eqref{eq:Psiproduct} leads to the same joint coordinate density as the (anti)symmetric state $\ket{\Psi_\rs}$ of Eq.~\eqref{eq:Psisym}.
Moreover, for non-overlapping wavefunctions, also one of the
terms on the r.h.s. of Eq.~\eqref{c2} vanishes. For example,
if $x_1$ and $x_2$ are chosen such that $|\psi_{L}(x_1)| \neq 0$ and $|\psi_{R}(x_2)| \neq 0$, then the second term in Eq.~\eqref{c2} vanishes, showing that there is no difference between using the symmetrized measurement operator in Eq.~\eqref{lalo2} or the non-symmetric operator $\hat{P}_{x_1 x_2}^{(\mr{ns})} = \ket{x_1} \bra{x_1} \otimes \ket{x_2} \bra{x_2}$. 

When, however, states $\ket{L}$ and $\ket{R}$ do overlap in space, i.e., $\braket{L | x} \braket{x | R} \not \simeq 0$ [in contrast to a possible $\braket{L | R} = \int \d x \, \psi_L^* (x) \psi_R(x) \simeq 0$], then all the three expressions are in general different, and $D_{\Psi}(x_1, x_2) = |\psi_L(x_1)|^2 \, |\psi_R(x_2)|^2$ obtained with the non-symmetric operator $\hat{P}_{x_1 x_2}^{(\mr{ns})}$ is ruled out for identical particles due to its manifest asymmetry.

\subsection{Simplest symmetric measurements}

Consider the simplest case of $N = 2$ in Eq.~\eqref{symproj} with the single-particle projectors $\hat{P}$ and $\hat{P}' = \id - \hat{P}$. The symmetric measurement on two identical particles of Eq.~\eqref{symproj} now has three projectors, 
\begin{equation} \label{777}
\hat{P} \otimes \hat{P}, \qquad \hat{P}' \otimes \hat{P}', \qquad \hat{P} \otimes \hat{P}' + \hat{P}' \otimes \hat{P} , 
\end{equation}
corresponding to the detection (i.e., measurement outcome $1$) of both, none, or one of the particles, respectively.
The measurement probabilities for states \eqref{eq:Psiproduct} and \eqref{eq:Psisym} are given by
\begin{eqnarray} \label{bebe1}
\braket{\Psi_\mr{p} |\hat{P} \otimes \hat{P}| \Psi_\mr{p} } &=& \braket{L | \hat{P} | L} \braket{R | \hat{P} | R},
\\ \nonumber
\braket{\Psi_{\mr{s}}|\hat{P} \otimes \hat{P}| \Psi_{\mr{s}} } &=& \braket{L R| \hat{P} \otimes \hat{P} | LR} 
+ \eta \braket{L R| \hat{P} \otimes \hat{P} |R L }
\\ \label{bebe2}
&=& \braket{L | \hat{P} | L} \braket{R | \hat{P} | R} 
+ \eta |\braket{L | \hat{P} | R}|^2, \hspace{-1.5 cm}
\end{eqnarray}
and the other probabilities $\braket{\hat{P}' \otimes \hat{P}'}$, $\braket{\hat{P} \otimes \hat{P}'}$ and 
$\braket{\hat{P}' \otimes \hat{P}}$ are easily obtained using the above equations. 

Note that the interference term $\propto \eta$ in Eq.~\eqref{bebe2}, which allows $\hat{P} \otimes \hat{P}$ to distinguish $\ket{\Psi_{\mr{s}}}$ from $\ket{\Psi_{\mr{p}}}$, will also appear when considering mixed states that contain both symmetrized and product states. As an example, consider a mixed state of the system given by the density operator 
\begin{equation}
\hat{\sigma} = (1 - p) \ket{\Psi_\mr{p}} \bra{\Psi_\mr{p}} + p \ket{\Psi_\mr{s}} \bra{\Psi_\mr{s}} , 
\label{eq:sigmamix}
\end{equation}
where $p$ denotes the probability that the state is symmetrized. Using the above relations, we obtain
\begin{equation}
\tr [ \hat{\sigma} \hat{P} \otimes \hat{P}] = \braket{L | \hat{P} | L} \, \braket{R | \hat{P} | R} 
+ p \eta |\braket{L | \hat{P} | R}|^2 . \label{eq:Trsig}
\end{equation}
This example will also be useful in Sec.~\ref{sec:sy}.

We may choose $\hat{P}$ to correspond to a particle detector of width $2 \ell$ placed at some position $\bar{x}$,
\begin{equation}
\hat{P} = \int_{\bar{x} - \ell}^{\bar{x} + \ell} \d x \, \ket{x} \bra{x} . \label{ell}
\end{equation}
Then the measurements described by Eqs.~\eqref{bebe1} and \eqref{bebe2} can reveal the difference between
$\ket{\Psi_\rp}$ and $\ket{\Psi_\rs}$ even for $\braket{L | R} = 0$, provided the two particle wavefunctions
overlap, $|\braket{L |x} \braket{x | R}| \neq 0$, for some coordinates $x$. 
This issue will be further discussed in Sec.~4\ref{sec:ada}.

\subsection{Measurements via coordinate superposition states}
\label{sec:m-p}

There is a class of measurements that can distinguish between the states $\ket{\Psi_\mr{p}}$ and $\ket{\Psi_\mr{s}}$ even for vanishing overlap $\braket{L | x} \braket{x | R} = 0 \, \forall \, x$.
Consider the projector 
\begin{equation}
\hat{P}_C = \ket{C} \bra{C}, \qquad \braket{C | C} = 1,
\end{equation}
where $\ket{C}$ involves a coherent superposition of coordinates, in contrast to $\hat{P}$ of Eq.~\eqref{ell}
which is an incoherent mixture of coordinate eigenstates. 

Let us assume that the single-particle wavepackets $\psi_{L, R}(x) = \braket{x |L, R}$ separated by distance $r$
are centered around positions $\mp r/2$ with the localization length $r_{L,R} \ll r$.
We take $\langle x|C\rangle\propto r^{-1/2}$ to be nearly constant for $x \in [-\frac{r}{2}, \frac{r}{2}]$
and quickly falling to zero for $|x| > \frac{r}{2}$, such that $\langle C|C\rangle=1$ is normalized. 
Substituting $\hat{P}_C$ into Eq.~\eqref{bebe2} we obtain 
\begin{equation} \label{pipo}
\braket{\Psi_\mr{s} | \hat{P}_C \otimes \hat{P}_C | \Psi_\mr{s}} = (1 + \eta) \, |\braket{R | C} \braket{C | L}|^2 
= (1 + \eta) \; O \Big( \frac{r_L r_R}{r^2}\Big),
\end{equation}
where $O$ corresponds to the standard asymptotic big-$O$ notation, while 
\begin{equation} \label{star}
\braket{C |L, R} \simeq \frac{1}{\sqrt{r}}\int_{-r/2}^{r/2}\d x\, \psi_{L,\,R}(x).
\end{equation}
The measurement probability in Eq.~\eqref{pipo} is small due to the
assumed $r \gg r_{L}, r_{R}$, but the correlation factor may still be
detectable. Thus, in the case of fermions ($\eta = -1$), we have for the antisymmetric state
$\braket{\Psi_\mr{s} | \hat{P}_C \otimes \hat{P}_C | \Psi_\mr{s}} = 0$
exactly, whereas for the product state $\braket{\Psi_\mr{p} | \hat{P}_C
\otimes \hat{P}_C | \Psi_\mr{p}}= O(r_L r_R / r^2)$. 

We can further increase the measurement probability of Eq.~\eqref{pipo} if we use for $\ket{C}$ a superposition 
of two states,
\begin{equation}
\ket{C} = \frac{1}{\sqrt{2}}(\ket{A_1} + \ket{A_2}),\label{zeno}
\end{equation}
with the condition $\braket{A_i | A_k} = \delta_{i k}$ (for $i, k = 1, 2$) to guarantee that 
$\braket{C | C} = 1$ is normalized and $\hat{P}_C^2 = \hat{P}_C$ is a projector. 
For simplicity, we also assume that $\braket{A_1 | R} = 0$ and $\braket{A_2 | L} = 0$.
With such a $\hat{P}_C = \ket{C} \bra{C}$, from Eqs.~\eqref{bebe1} and \eqref{bebe2} we obtain
\begin{eqnarray}
\label{bebe4}
\braket{\Psi_\mr{p} | \hat{P}_C \otimes \hat{P}_C | \Psi_\mr{p}} &=& \frac{1}{4} |\braket{A_1 | L}|^2 |\braket{A_2 | R}|^2, \\
\label{bebe5}
\braket{\Psi_\mr{s} | \hat{P}_C \otimes \hat{P}_C | \Psi_\mr{s}} &=&
\frac{1}{4} (1 + \eta) |\braket{A_1 | L}|^2 |\braket{A_2 | R}|^2.
\end{eqnarray}
Now, if $|\braket{A_1 | L} \braket{A_2 | R}| \sim 1$, the measurement probabilities of Eqs.~\eqref{bebe4} and \eqref{bebe5} will be sizable. Hence, the interference term $\propto \eta$ in Eq.~\eqref{bebe5} will contribute to the measurement probabilities, which can therefore reveal the difference between $\ket{\Psi_\mr{p}}$ and $\ket{\Psi_\mr{s}}$. We emphasize again that these are valid measurements even for vanishing spatial overlap $\braket{L | x} \braket{x | R} \simeq 0 \, \forall \, x$, but such measurements will be impossible to realize when the particles are very far apart, since in the scenario of Eq.~\eqref{star} $\braket{C |L, R} \to 0$, and in the scenario of Eq.~\eqref{zeno} it will be difficult to realize a projector onto a superposition of well-separated states \cite{Frowis2018}. Other complications may arise from considering characteristic times of such measurements that are not instantaneous, which is, however, beyond the scope of this work.

\section{Symmetrization of a two-particle state}
\label{sec:sy}

We now turn to the dynamics of the system. We assume that at time $t = 0$ two identical particles, $1$ and $2$, are produced by two different spatially well-separated sources $\mc{S}_L$ and $\mc{S}_R$ in states $\ket{L}$ and $\ket{R}$, respectively. Due to a large interparticle distance, their wavefunctions $\psi_{L,R} (x,t) = \braket{x | L(t), R(t)}$ have vanishing initial overlap $\psi_{L}^* (x,0) \psi_{R} (x,0) = 0 \; \forall \, x$ and hence $\braket{L | R} = 0$. Although the particles are identical, they are distinguishable at early times by the very fact of being produced far apart from each other: the particle labels $1$ and $2$ are meaningful and remain so at least for some time. We thus adopt the initial product state hypothesis and write the two particle state as a product state $\ket{\Psi(t)} = \ket{\Psi_\mr{p}(t)} = \ket{L(t)} \otimes \ket{R(t)}$ as per Eq.~\eqref{eq:Psiproduct}. We furthermore assume that the particles do not interact unless they are at the same location. Until then, the state of each particle evolves independently and their individual dynamics are governed by the same single-particle Hamiltonian $H$, leading to
\begin{equation} \label{kara}
\ket{L(t)} = e^{-i \hat{H} t / \hbar} \ket{L}, \qquad \ket{R(t)} = e^{-i \hat{H} t / \hbar} \ket{R}.
\end{equation}
Hence, the state overlap is conserved in time, $\braket{L(t) | R(t)} = \braket{L | R} = 0$.
Yet, as the particles approach each other, their probability densities start to overlap: 
$|\psi_{L} (x,t)|^2 |\psi_{R} (x,t)|^2 \neq 0$ for some interval of values of $x$. 
Eventually, the particles collide at some time $t_c$, which erases their distinguishability.
This amounts to a transition to the symmetrized state
$\ket{\Psi(t)} = \ket{\Psi_{\mr{s}}(t)} = \frac{1}{\sqrt{2}}(\ket{L(t)} \otimes \ket{R(t)} 
+ \eta \ket{R(t)} \otimes \ket{L(t)})$ of Eq.~\eqref{eq:Psisym}, with $\eta = \pm 1$ for bosons/fermions. 

The collision time $t_c$ is a random variable, with the probability density $\rho(t_c)$ determined in Sec.~\ref{sec:tc}. We describe the transition from product state $\ket{\Psi(t < t_c)} = \ket{\Psi_{\mathrm{p}}(t)}$ to symmetrized state $\ket{\Psi(t > t_c)} = \ket{\Psi_{\mathrm{s}}(t)}$ by a quantum jump at time $t = t_c$ realized by the projector $\hat{\Pi} = \frac{1}{2} (1 + \eta \hat{\pi})$, with $\hat{\pi}$ being the permutation operator:
\begin{equation} \label{eq:qjump}
\ket{\Psi_{\mathrm{p}}} \to \frac{\hat{\Pi} \ket{\Psi_{\mathrm{p}}}}
{\sqrt{\bra{\Psi_{\mathrm{p}}} \hat{\Pi} \ket{\Psi_{\mathrm{p}}}}} = \ket{\Psi_{\mathrm{s}}}.
\end{equation} 
Note that the permutation operator $\hat{\pi}$ commutes with any Hamiltonian for identical particles. Hence, once the state is symmetrized upon particle collision, it will no longer change by the subsequent application of the projector $\hat{\Pi}$.

Even though we deal with a closed, non-dissipative system, the spontaneous symmetrization of Eq.~(\ref{eq:qjump}) is formally identical to a quantum jump in a continuously monitored dissipative system \cite{Dalibard1992, Dum1992, Gardiner1992, Plenio1998}. In a usual quantum measurement, the experimenter is free to choose the time at which the measurement is performed. Such a measurement is realized by coupling the system to a measuring apparatus which disrupts the coherent evolution of the closed system for a certain period of time, controlled by the experimenter, during which the measurement takes place. When measuring time, however, the time of registering the measurement outcome is random and coincides with the measurement outcome itself. The measurement is realized by the collision itself, and takes place without the presence of an external apparatus, i.e., the state of the system becomes symmetrized whether or not we query the transition time.

Thus, in a single realization, or quantum trajectory, the state of a pair of particles evolves according to the random but pure state
\begin{equation}
\ket{\Psi(t)} = \theta(t_c-t) \ket{\Psi_{\mathrm{p}}(t)} 
+ \theta(t-t_c) \ket{\Psi_{\mathrm{s}}(t)}, \label{kora}
\end{equation}
where $\theta(t)$ is Heaviside step function. 
For an ensemble of particles prepared in state $\ket{\Psi_{\mathrm{p}}(0)}$ at time $t = 0$, 
the probability of transition \eqref{eq:qjump} to happen during the time interval $[0, t]$ is 
\begin{equation}
p_c(t) = \int_0^t \d t_c \, \rho(t_c) . \label{eq:pc} 
\end{equation} 
Hence, the ensemble-averaged density operator for the system is given by
\begin{equation} 
\hat{\sigma}(t) = 
\big( 1-p_c(t) \big) \ket{\Psi_{\mr{p}}(t)} \bra{\Psi_{\mr{p}}(t)} + p_c(t) \ket{\Psi_{\mr{s}}(t)} \bra{\Psi_{\mr{s}}(t)} . 
\label{eq:sigmat} 
\end{equation} 
Now, whenever $p_{c} (t) \to 1$, the symmetrization is complete, $\hat{\sigma} = \ket{\Psi_{\mr{s}}} \bra{\Psi_{\mr{s}}}$;
if $p_{c}(t)=0$ for all times, as for initially separated particles propagating away from each other, 
the state remains a tensor product, $\hat{\sigma} = \ket{\Psi_{\mr{p}}} \bra{\Psi_{\mr{p}}}$;
while in general, $0 < p_c < 1$, we have a mixed state of Eq.~\eqref{eq:sigmat}. 

Single quantum trajectories and the ensemble-averaged evolution of the system can also be simulated numerically using the stochastic wavefunction approach \cite{Dalibard1992, Dum1992, Gardiner1992, Plenio1998}. Namely, to simulate a quantum trajectory (a single realization of experiment) starting from the product state $\ket{\Psi(t=0)} = \ket{\Psi_{\mr{p}}}$, we draw from a uniform distribution a random number $\gamma \in [0,1]$ and compare it with the collision probability of Eq.~\eqref{eq:pc}. The transition \eqref{eq:qjump} from the product state $\ket{\Psi_{\mr{p}}}$ to the symmetrized state $\ket{\Psi_{\mr{s}}}$ occurs at time $t_c'$ when $p_c(t_c') = \gamma$. Subsequently, the two-particle state continues to evolve as a symmetrized state, $\ket{\Psi(t > t_c')} = \ket{\Psi_{\mr{s}}}$. The density operator of the system, $\hat{\sigma}(t)$, is obtained by averaging over $M \gg 1$ independently simulated trajectories $\ket{\Psi^{(m)}(t)}$: 
\begin{equation}
\hat{\sigma}(t) = \lim_{M \to \infty} \frac{1}{M} \sum_{m=1}^M
\ket{\Psi^{(m)}(t)}\bra{\Psi^{(m)}(t)} . 
\end{equation}

\section{Collision time of two particles}
\label{sec:tc}

Our next task is to determine the probability density of the collision times $\rho(t_c)$, given the initial states of the particles each evolving under the free Hamiltonian. In subsection \ref{sb} we derive $\rho(t_c)$ using a quasi-classical approach, while in subsection \ref{sc} we present the fully quantum treatment, followed by the numerical results in subsections \ref{sec:2pGauss} and \ref{sec:ada}. 

\subsection{Single-particle dynamics}
\label{sec:sa}

The Schr\"{o}dinger equation for a single-particle wavefunction evolving under the free propagation Hamiltonian $\hat{H} = \frac{1}{2 m} \hat{p}^2$ is easily solved in the momentum representation, $i\partial_t \psi(p,t)=\frac{p^2}{2m}\psi(p,t)$ ($\hbar = 1$), leading to
\begin{equation}
\psi(p,t)=e^{-itp^2/(2m)}\chi(p), \label{eq:psipt}
\end{equation}
where $\chi(p)$ is a normalized function, $\int \d p |\chi(p)|^2 = 1$. In subsection \ref{sb} we discuss in the quasi-classical regime with non-negative Wigner functions corresponding to the single-particle wavefunctions \cite{Hudson1974}. We therefore consider Gaussian wavefunctions $\psi_{L,R}(p, t) = \braket{p | L(t),R(t)}$ and take $|\chi(p)|^2$ to be a normalized Gaussian function
\begin{eqnarray} \label{eq:chip}
\chi(p) &=& \frac{\sqrt{a}}{\pi^{\frac{1}{4}}} \exp \left[-\frac{b^2}{2}+p(ba-ic)-\frac{a^2p^2}{2} \right],
\end{eqnarray}
with constants $a>0$ and $b$ defined via the mean and variance of the momentum through
\begin{equation} \label{gustaf}
\langle p\rangle\equiv\int\d p\,p|\chi(p)|^2=\frac{b}{a}, \qquad \langle p^2\rangle-\langle p\rangle^2=\frac{1}{2a^2},
\end{equation}
while $c$ is the initial center of mass coordinate of the wavepacket. Recalling the $\delta$-function normalized momentum eigenfunctions, $\braket{x | p} = \frac{1}{\sqrt{2 \pi}}\, e^{i p x}$, we obtain from Eq.~\eqref{eq:chip} the wavefunction in the coordinate representation:
\beaa \label{eq:psixt}
\psi(x, t) &= \frac{1}{\sqrt{2 \pi}} \int \d p \, e^{ipx}\psi(p,t)
= \frac{1}{\pi^{\frac{1}{4}}\sqrt{a (1 + i \tau)}} \exp\left[-\frac{b^2}{2} - \frac{(\xi - i b)^2}{2(1 + i \tau)} \right]
\\
&= \frac{1}{\pi^{\frac{1}{4}}\sqrt{a (1 + i \tau)}} \exp\left[-\frac{(\xi-b\tau)^2}{2(1+\tau^2)} + \frac{i}{2}\,\,\frac{2b\xi+\tau(\xi^2-b^2)}{1+\tau^2} \right], ~~
\eeaa
where $\xi \equiv (x-c)/a$ and $\tau \equiv t/(ma^2)$ are the dimensionless coordinate and time, respectively.

As the particles approach each other, eventually their spatial probability densities start to overlap. 
Yet, their wavefunction overlap is conserved in time, $\langle L(t)|R(t) \rangle = \langle L|R \rangle$, 
and remains small if initially $\langle L|R \rangle \ll 1$. 
Indeed, using Eq.~\eqref{eq:chip}, we can verify that the overlap is given by 
\beaa \label{eq:ovrlpLR}
\braket{L|R} &= \int \d p \,\chi_L^*(p)\chi_R(p)
\\
&= \sqrt{\frac{2a_L a_R}{a^2_L+a^2_R}}
\exp \left[-\frac{(a_Lb_R-a_Rb_L)^2 + (c_R-c_L)^2}{2(a^2_L+a^2_R)} + i \frac{(c_L-c_R)(a_Rb_R+a_Lb_L)}{a^2_L+a^2_R} \right], ~~
\eeaa
where $a_{L, R}$, $b_{L, R}$, and $c_{L, R}$ are the parameters of the Gaussian wavepackets in Eq.~\eqref{eq:chip}.
We thus see that $|\langle L|R\rangle| \ll 1$ either for initially largely separated particles, $\frac{(c_R-c_L)^2}{2(a^2_L+a^2_R)}\gg 1$, or for particles moving in the opposite directions, $\frac{(a_Lb_R-a_Rb_L)^2}{2(a^2_L+a^2_R)}\gg 1$.

\subsection{Quasi-classical collision time}
\label{sb}

Consider two classical point-like particles with trajectories $x_{1,2}(t) = x_{1,2}(0) + \frac{p_{1,2}(0)}{m}t$. The particles collide at time 
\begin{equation}
t_c = m \frac{x_1(0) - x_2(0)}{p_2(0) - p_1(0)} , \label{eq:tcclass}
\end{equation}
which makes them practically indistinguishable thereafter. 
Given now the initial factorized distributions $D_L(x,p)$ and $D_R(x,p)$ of coordinates and momenta of the particles 1 and 2, the probability density of collision times $t_c$ is given by
\begin{eqnarray}\nonumber
\rho_{\mathrm{cl}}(t_c) & = & \iiiint \d x_1 \, \d p_1 \, \d x_2 \, \d p_2 \, 
\delta \left(t_c - m \frac{x_1 - x_2}{p_2 - p_1} \right) D_L(x_1, p_1) D_R(x_2, p_2) 
\\ \label{eq:rhocl}
& = & \iiint \d p_1 \, \d x_2 \, \d p_2 \, \frac{|p_1-p_2|}{m} D_L 
\left(x_2 - \frac{p_1 - p_2}{m} t_c, p_1 \right) D_R(x_2, p_2), 
\end{eqnarray}
and is normalized as $\int \d t_c \, \rho_\mathrm{cl}(t_c) = 1$ provided that 
both $D_{L,R}(x,p)$ are normalized. 

To describe the collision of quantum particles, we can use in the above equation 
instead of $D_{L,R}(x,p)$ an appropriate quasi-probability distribution.
A well-behaved (non-negative) quasi-probability distribution for Gaussian wavefunctions 
of the particles is the Wigner function \cite{Peres,Hudson1974} 
\[
W(x,p,t) = \frac{1}{\pi} \int \d q \, \psi^*(p+q,t) \psi(p-q,t)\,e^{-2ixq} .
\]
Using the wavefunctions $\psi(p,t)$ of Eq.~\eqref{eq:psipt} at time $t=0$, we then obtain 
the probability density of collision times $t_c$,
\beaa \label{eq:rhotcsc}
\rho_{\mathrm{cl}}(t_c) &= \frac{1}{\pi} \iiint \d q \, \d p_1 \d p_2 \frac{|p_1-p_2|}{m} 
\chi^{*}_L(p_1 + q) \, \chi_L(p_1-q) \, 
\chi^{*}_R(p_2 - q) \, \chi_R(p_2 + q) \, e^{2iq t_c \frac{p_1 - p_2}{m}}
\\ 
&= \frac{1}{\pi m}\,\frac{a_R a_L}{a_R^2+a_L^2} \int\d u\, |u| \,
\exp \! \Bigg[\! -\frac{a^2_R a^2_L \big(u+ \frac{b_R}{a_R} - \frac{b_L}{a_L}\big)^2}{a_R^2 + a_L^2} \Bigg]
\exp \! \Bigg[\! -\frac{\big(\frac{t_c u}{m} + c_L -c_R\big)^2}{a_R^2 + a_L^2} \Bigg].
\eeaa
This integral over the relative momentum $u \equiv p_2 - p_1$ can be expressed through the error function.
Note that the first Gaussian in the second line of Eq.~\eqref{eq:rhotcsc} is centered at momentum difference $u \simeq \frac{b_L}{a_L}-\frac{b_R}{a_R}$, 
while the second Gaussian is centered around the initial interparticle distance $u t_c/m \simeq (c_R-c_L)$.

\subsection{Quantum collision time}
\label{sc}

We now present a fully quantum derivation of the collision time, following the ideas
of Refs.~\cite{Aharonov1961, Grot1996, Muga1998} on the quantum arrival time.
As a quantum generalization of the classical collision time in Eq.~\eqref{eq:tcclass}, we define a collision time operator 
\bea \label{eq:tcquant}
\hat{t}_c = - \frac{m}{2} \bigg\{\hat{x} \otimes \id - \id \otimes \hat{x}, \frac{1}{\hat{p} \otimes \id - \id \otimes \hat{p}} \bigg\},
\eea
where $\{\; , \; \}$ is the anticommutator. Note that $\hat{t}_c$
is symmetric with respect to the particle permutation: $\hat{\pi} \, \hat{t}_c \hat{\pi} = \hat{t}_c$.

In the momentum representation, denoting $\hat{p} \otimes \id = \hat{p}_1$, $\id \otimes \hat{p} = \hat{p}_2$, 
we have $\hat{x} \otimes \id = i \partial_{p_1}$, $\id \otimes \hat{x} = i \partial_{p_2}$. 
Introducing the relative and center of mass momenta, $u = p_2 - p_1$ and $v = \frac{1}{2} (p_1 + p_2)$, 
we have for any state $\ket{\phi}$ 
\begin{equation} \label{eq:tcqu}
\braket{u, v | \hat{t}_c | \phi} = - {i m} \bigg\{\frac{1}{u}, \partial_u \bigg\} \phi(u, v).
\end{equation}
This expression does not depend on $v$ due to translational invariance.
Denoting the tensor product of $u$ space and $v$ space by $\circ$ (reserving $\otimes$ for the tensor product of the original single-particle Hilbert spaces), we can recast Eq.~\eqref{eq:tcqu} as
\begin{equation} \label{eq:tcfromt}
\hat{t}_c = \hat{t} \circ \id^{(v)},
\end{equation}
where $\id^{(v)}$ is the identity operator in the $v$ space, and 
\begin{equation} \label{eq:tquant}
\hat{t} = - {i m} \bigg\{\frac{1}{u}, \partial_u \bigg\} = - i m \bigg(\frac{2}{u} \partial_u - \frac{1}{u^2}\bigg)
\end{equation}
is precisely the quantum time-of-arrival operator \cite{Aharonov1961, Grot1996, Muga1998, Mugaetal}
that lives in the $u$ space.

Due to singularity of $\hat{t}$ at $u = 0$, for each eigenvalue $t_c$, there are two linearly independent eigenvectors $\ket{t_c, +}$ and $\ket{t_c, -}$ which we choose as \cite{Egusquiza1999}
\begin{equation} \label{rok}
\braket{u | t_c, \pm} = \frac{\theta(\pm u)}{\sqrt{4 \pi m}} \sqrt{|u|} \, e^{i t_c u^2/(4m)};
\end{equation}
see Appendix~\ref{app:degeneracy} for a detailed discussion. The functions $\braket{u | t_c, +}$ and $\braket{u | t_c, -}$ correspond, respectively, to relative momenta $u < 0$ and $u > 0$. Note that, although we start counting from $t = 0$, $t_c$ ranges from $-\infty$ to $\infty$ to account for both future collisions of the particles moving towards each other, and (possible) past collisions of the particles that are moving away from each other. The eigenprojector of $\hat{t}$ corresponding to the eigenvalue $t_c$ is then
\begin{equation} \label{baikonur}
\hat{Y}(t_c) = \ket{t_c, +} \bra{t_c, +} \, + \, \ket{t_c, -} \bra{t_c, -}.
\end{equation}
The eigenprojectors are complete in the $u$ space: $\int \d t_c \hat{Y}(t_c) = \id^{(u)}$. 
The eigenprojector of $\hat{t}_c $ corresponding to the eigenvalue $t_c$ can be written as
\begin{equation} \label{buran}
\hat{P}(t_c) = \hat{Y}(t_c) \circ \id^{(v)},
\end{equation}
and we immediately see that
\begin{gather} \label{bronto}
\int \d t_c \hat{P}(t_c) = \id^{(u)} \circ \id^{(v)} = \id \otimes \id,
\\ \label{eq:Ptc}
\int \d t_c \bra{p'_1, p'_2} \hat{P}(t_c) \ket{p_1, p_2} = \delta(p_1 -p'_1) \, \delta(p_2 -p'_2) .
\end{gather}

Importantly, even though $\hat{t}_c$ is formally Hermitian and its eigenprojectors satisfy the completeness relation \eqref{bronto}, 
it is \textit{not} a self-adjoint operator \cite{Paul1961, Allcock1969, Gieres2000}
\footnote{Although this operator is not self-adjoint, it can be made Hermitian (also referred to as ``symmetric'' in the mathematical literature) in a suitable Hilbert space; see Ref.~\cite{Gieres2000} for an accessible review on the difference between Hermitian and self-adjoint operators.} 
and the eigenvectors of $\hat{t}$ (and therefore of $\hat{t}_c $) are not orthogonal:
\begin{equation} \label{nigeria}
\braket{t_1, \pm | t_2, \pm} = \frac{1}{2 \pi} \int_0^\infty \d u \, e^{i u (t_2 - t_1 + i 0)} = \frac{1}{2 \pi} \, \mathcal{P} \bigg[\frac{1}{t_1 - t_2}\bigg] + \frac{i}{2} \delta(t_1 - t_2),
\end{equation}
where we used Eq.~\eqref{rok} and the Sokhotski-Plemelj theorem, and $\mathcal{P}$ denotes the principal value map. That $\hat{t}_c$ is not a self-adjoint operator is a manifestation of the fact that there can be no time observable in quantum mechanics \cite{Allcock1969, Egusquiza1999, Mugaetal, Leon2017} (see also Appendix~\ref{app:time}).

Even though the eigenresolution of $\hat{t}_c $ does not define a projective measurement due to Eq.~\eqref{nigeria}, the fact that operators $\ket{t_c, \pm} \bra{t_c, \pm} \circ \id^{(v)}$ are Hermitian, positive-semidefinite, and sum up to the identity, as per Eqs.~\eqref{baikonur}--\eqref{bronto}, means that they constitute a POVM \cite{Peres}. Hence, we can adopt the view that the measurement of $\hat{t}$, and thus $\hat{t}_c$, realizes the POVM defined by its eigenresolution \cite{Giannitrapani1997, Muga1998, Mugaetal}. An important difference between the usual projective measurements and general POVMs is that, for the latter, there is no unique prescription for identifying the post-measurement states, i.e., there is unitary freedom. This is also true for the measurement of $\hat{t}_c$. As explained in Appendix \ref{app:povm}, this freedom can be used to choose the post-measurement state to be $\ket{\Psi_{\rs} (t_c)}$, so that the symmetrization jump in Eq.~\eqref{eq:qjump} can be thought of as the post-measurement state after the collision time $\hat{t}_c$ is measured.

Using the two-particle \textit{product} state in Eq.~\eqref{eq:Psiproduct} in the momentum representation, $\braket{p_1,p_2|\Psi(t)} = \psi_L(p_1,t) \, \psi_R(p_2,t)$, the probability density of the collision times is given by
\beaa \label{eq:ghana}
\rho_t(t_c) &= \bra{\Psi(t)}\hat{P}(t_c) \ket{\Psi(t)}
\\
&= \iiiint \d p_1 \d p_2 \, \d p'_1 \d p'_2 \, \braket{\Psi(t) | p'_1, p'_2} \bra{p'_1, p'_2} \hat{P}(t_c) \ket{p_1, p_2} \braket{p_1, p_2 | \Psi(t)}
\\
&= \sum_{s = \pm} \iiint \d v \, \d u \, \d u' \, \psi_L^*\Big(v - \frac{u'}{2}, t\Big) \psi_R^*\Big(v + \frac{u'}{2}, t\Big) \psi_L\Big(v - \frac{u}{2}, t\Big) \psi_R\Big(v + \frac{u}{2}, t\Big) \braket{u' | \hat{Y}(t_c) | u} .
\eeaa
Consistently with the discussion above [see Eq.~\eqref{pauli1}], the collision time distribution depends on the initial time $t$ as $\rho_t(t_c) = \rho_{t = 0}(t_c + t)$: if the system is prepared in state $\ket{\Psi(t)}$ at time, say, $t > 0$ instead of $t = 0$, the collision will happen earlier.

\subsection{Gaussian wavepackets} 
\label{sec:2pGauss}

We now assume that two Gaussian wavepackets described by Eq.~\eqref{eq:psipt}, initially (at $t = 0$) separated by a distance $d \equiv c_R - c_L > 0$, are moving towards each other. Omitting the subscript of $\rho$ in Eq.~\eqref{eq:ghana} ($\rho_{t = 0} \to \rho$), we have 
\beaa \label{eq:rhotcq} 
\rho(t_c) =& \frac{1}{2\pi^{3/2} m}  \frac{a_L a_R}{\sqrt{a^2_L+a^2_R}} 
\exp\left[-\frac{(a_Lb_R-a_Rb_L)^2}{a^2_L+a^2_R}\right] 
\\
& \times \! \iint_0^\infty \! \d u \, \d u' \, \sqrt{uu'} \, \exp \! \Big[\frac{it_c}{4 m} (u'^2 - u^2) - A_2 (u'^2 + u^2)+ A_1 u u'\Big] \cosh \! \Big[\frac{i}{2}(c_R-c_L)(u'-u)+A_0(u+u') \Big], ~~~
\eeaa
where 
\bea \nonumber
A_0 = \frac{a_L a_R (a_L b_R - a_R b_L)}{a_L^2 + a_R^2}, \qquad A_1 = \frac{(a_R^2 - a_L^2)^2}{8 (a_L^2 + a_R^2)}, \qquad A_2 = \frac{a_R^4 + 6 a_R^2 a_L^2 + a_L^4}{16 (a_L^2 + a_R^2)}.
\eea

\begin{figure}[t]
\subfloat[]{\includegraphics[width = 0.49 \textwidth]{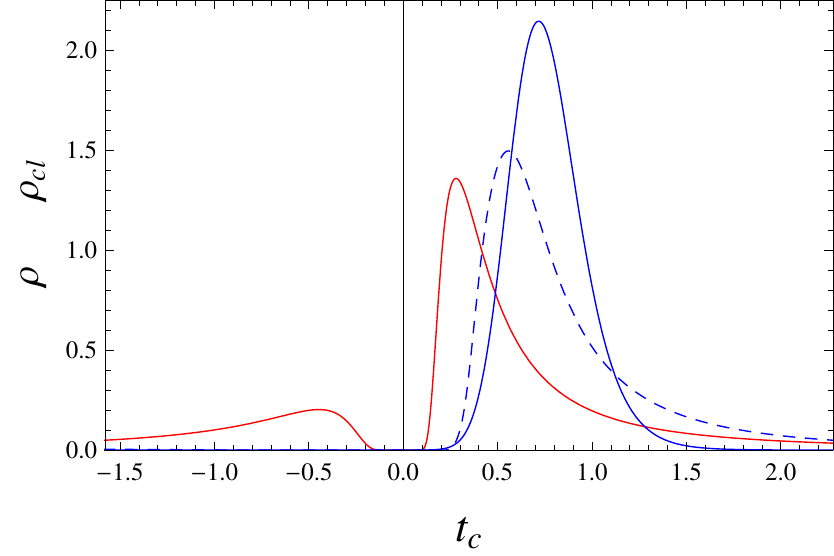}}
\subfloat[]{\includegraphics[width = 0.49 \textwidth]{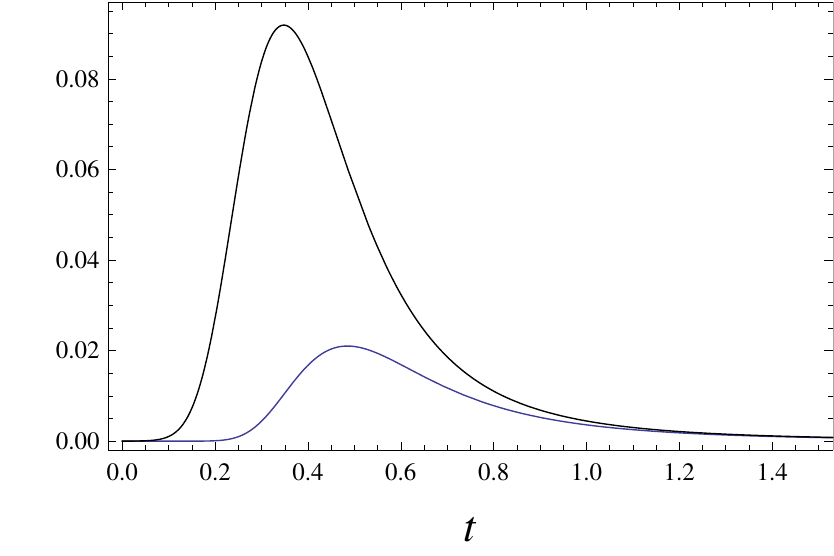}}
\caption{(a) Probability densities $\rho(t_c)$ and $\rho_{\mr{cl}}(t_c)$ of the collision time $t_c$, as obtained from Eqs.~\eqref{eq:rhotcsc} and \eqref{eq:rhotcqsmp} for two Gaussian wavepackets with large initial separation $c_R - c_L = 5$, and $a_{R, L} = 1$, $b_L = -b_R = 3.33$ (blue solid line); $a_{R, L} = 0.3$, $b_L = -b_R = 1$ (blue dashed line); and $a_{R, L} = 0.1$, $b_L = -b_R = 0.333$ (red solid line). For all curves the difference between the quantum $\rho(t_c)$ and semiclassical $\rho_{\mr{cl}}(\tau)$ probability densities is indiscernible. 
(b) The magnitudes of the first (black line) and second (blue line) terms on the r.h.s. of Eq.~\eqref{eq:Trsigt}, for a pair of Gaussian wavepackets with $c_R - c_L = 2.5$, $a_{R, L} = 0.5$, and $b_L = -b_R = 1.5$, as measured by a coordinate detector of Eq.~\eqref{ell} placed in the two particle collision region, $\bar{x} = (c_L + c_R)/2$ and $\ell = 0.25$. The collision probability $p_c(t)$ of Eq.~\eqref{eq:pc} with $\rho(t_c)$ of Eq.~\eqref{eq:rhotcqsmp} is well localized in the interval of times $t \in [0.1, 2]$. We have set $m = 1$ and $\hbar = 1$ in both figures.}
\label{fig:rhocol_trsig}
\end{figure}

For $a = a_L = a_R$, Eq.~\eqref{eq:rhotcq} simplifies to 
\beaa \label{eq:rhotcqsmp}
\rho(t_c) = \frac{a \, e^{- \frac{1}{2}(b_R - b_L)^2}}{4 \pi m \sqrt{2 \pi}} \Bigg(& \, \bigg| \int_0^\infty \d u \sqrt{u} \, \exp \bigg[\frac{u^2}{4} \Big(\frac{i t_\mr{c}}{m} - a^2\Big) + \frac{i u}{2} d + \frac{a u}{2}(b_R - b_L) \bigg] \bigg|^2
\\
&+ \bigg| \int_0^\infty \d u \sqrt{u} \, \exp \bigg[\frac{u^2}{4} \Big(\frac{i t_\mr{c}}{m} - a^2\Big) - \frac{i u}{2} d - \frac{a u}{2} (b_R - b_L)\bigg] \bigg|^2 \, \Bigg). ~
\eeaa
In Fig.~\ref{fig:rhocol_trsig}(a) we show the probability densities $\rho_{\mathrm{cl}}(t_c)$ and $\rho(t_c)$ 
of the collision time obtained from the semiclassical and fully quantum treatments, which are
nearly identical for initially separated Gaussian wavepackets with vanishing overlap $|\braket{L|R}| \simeq 0$.
There, the blue dashed and solid lines correspond to the particles with the same average 
momentum difference $\frac{b_L}{a_L}-\frac{b_R}{a_R}$, but for the former case the momentum uncertainty 
$\frac{1}{\sqrt{2} a_{L,R}}$ is larger, and therefore the dashed curve is lower and more smeared.
For the red curve, the average momentum is again the same, but the momentum uncertainty 
is even larger than the mean. Hence $\rho(t_c) \simeq \rho_{\mathrm{cl}}(t_c)$ has nonvanishing 
values even for negative times $t_c <0$, while the integrated collision probability 
$p_c \simeq \int_0^\infty \d t_c \, \rho(t_c) = 0.7469$ is smaller than one. 
This means that there is a finite probability $1-p_c \gtrsim 0.25$, given by Eq.~\eqref{eq:pc}, 
that the two particles move away from each other and never collide, 
i.e., the state $\hat{\sigma}(t)$ of Eq.~\eqref{eq:sigmat} remains mixed even for large times. 

\subsection{Coordinate detection} 
\label{sec:ada}

We finally illustrate the dynamics of the measurement probability \eqref{eq:Trsig}, namely
\begin{eqnarray} \label{eq:Trsigt}
\tr [ \hat{\sigma}(t) \hat{P} \otimes \hat{P}] = \braket{L(t) | \hat{P} | L(t)} \, \braket{R(t) | \hat{P} | R(t)} + \eta p_c(t) |\braket{R(t) | \hat{P} | L(t)}|^2 ,
\end{eqnarray}
assuming that $\ket{L(t)}$ and $\ket{R(t)}$ are identical, initially well-separated Gaussian wavepackets propagating 
toward each other, while the coordinate detector $\hat{P}$ of Eq.~\eqref{ell} is placed in the collision region.
Mathematically this is expressed by the fact that $\rho(t_c)$ is peaked around times $t$ when both
$\braket{L(t) | \hat{P} | L(t)} \braket{R(t) | \hat{P} | R(t)}$ and $p_c(t) |\braket{R(t) | \hat{P} | L(t)}|^2$
attain their maxima (see Appendix~\ref{app:Gauss} for analytical expressions for these quantities).
In Fig.~\ref{fig:rhocol_trsig}(b) we show the dynamics of the two terms on the r.h.s. of Eq.~\eqref{eq:Trsigt}.
As expected, the first term is several times larger than the second, correlation term, which, however, is still well pronounced.  
Note that a proper choice of the detector width $\ell$ in Eq.~\eqref{ell} is important for this conclusion, 
since for $\ell\to 0$ the detector reduces to a point and produces no signal, while for $\ell\to \infty$ we also have
$|\braket{R(t) | \hat{P} | L(t)}|^2 = |\braket{R(t) | L(t)}|^2 = |\braket{R(0) | L(0)}|^2 = 0$.

\section{Conclusions}
\label{sec:concl}

In summary, we have proposed a self-consistent model for the dynamics of identical particles produced independently by spatially separated sources and therefore described initially by a product state. The symmetrization of the two-particle state occurs dynamically as the particles collide with each other which erases their distinguishability. Our simple two-particle 1D model provides an intuitively plausible symmetrization picture. We show that, for Gaussian wavepackets, the collision probabilities can be accurately described via a quasi-classical approach employing Wigner functions of the particles. Quantum mechanically, the transition between the product and symmetric states can be formally regarded as a consequence of spontaneous measurement of the (non-self-adjoint) collision time operator. Semiclassical extension of the model to $3$D collisions of short-range interacting particles is straightforward, as discussed in Appendix~\ref{app:3D}, but a fully quantum generalization of the collision time operator to $3$D space is a known problem and will require further research.

Our model is consistent with the bulk of experimental observations of the (anti)symmetrization effects, such as, e.g., the Pauli principle or Bose-Einstein condensation, which consider identical particles that had sufficient time to interact and collide with each other. For such particles, we can safely assume that the symmetrization transition had already taken place before the system became subject to interrogation. The main purpose of our model is to provide a clear physical picture of the transient regime between the particles being very far (when their state is still a tensor product) and sufficiently close (when their state is already symmetrized), by employing several physically plausible assumptions. 

It is worth emphasizing that our results are not interpretational---they can be tested experimentally. Once such tests are preformed, the hypothesis that independently generated, spatially separated identical particles can be considered distinguishable can be supported or laid to rest.

\acknowledgements
We thank Janet Anders for interesting discussions on identical particles, and Roger Balian and Theo Neuwenhuizen for useful remarks. A.E.A. thanks Theo Neuwenhuizen for his role in motivating this research.
This work was supported by the SCS of Armenia, grants No. 18T-1C090 and No. 20TTAT-QTa003.


\appendix

\section{Reduced states of identical particles}
\label{app:margo}

Reduced (sometimes also called ``marginal'') states of multipartite quantum systems are defined through local observables. Consider a compound system in a Hilbert space $\mathcal{H} = \mathcal{H}_1 \otimes \mathcal{H}_2$ consisting of two
distinguishable subsystems $1$ and $2$ living, respectively, in $\mathcal{H}_1$ and $\mathcal{H}_2$.
Given a state of the total system, $\hat{\sigma}$, the reduced state of, say, the first subsystem is defined as a positive-semidefinite operator $\hat{\sigma}_1$ with $\tr \hat{\sigma}_1 = 1$ such that
\begin{equation} \label{marg}
\tr(\hat{\sigma}_1 \hat{O}) = \tr(\hat{\sigma} \, \hat{O} \otimes \id_2) \quad \forall \; \hat{O} \in \mathcal{L}(\mathcal{H}_1),
\end{equation}
where $\mathcal{L}(\mathcal{H}_1)$ is the algebra of observables $\hat{O}$ on $\mathcal{H}_1$ and $\id_2$ is the identity operator in $\mathcal{H}_2$.

For identical particles, the set of observables is restricted to permutation-symmetric operators, i.e., only such operators can be measured. This in turn implies that $\tr(\hat{\sigma} \, \hat{O} \otimes \id_2)$ is devoid of any physical meaning since any apparatus measuring a single-particle observable $\hat{O}$ in fact measures
$\hat{O} \otimes \id + \id \otimes \hat{O}$; in other words, it measures $\hat{O}$ for both particles.
Hence, the single-particle states are defined through
\begin{equation} \label{margsym}
\tr(\hat{\sigma}_1 \hat{O}) + \tr(\hat{\sigma}_2 \hat{O}) = \tr(\hat{\sigma} \, [\hat{O} \otimes \id + \id \otimes \hat{O}]) \quad \forall \; \hat{O} \in \mathcal{L}(\mathcal{H}_1) \cong \mathcal{L}(\mathcal{H}_2),
\end{equation}
where $\mathcal{H}_1 \cong \mathcal{H}_2$ for identical particles (the symbol $\cong$ means isomorphic).

When $\hat{\sigma}$ is permutation-symmetric, the standard approach is to require that $\hat{\sigma}_1 = \hat{\sigma}_2$, in which case Eq.~\eqref{margsym} uniquely determines the reduced state. For example, if the joint state is $\propto \ket{L} \otimes \ket{R} + \eta \ket{R} \otimes \ket{L}$, with some $\braket{L | R} = 0$ and $\eta = \pm 1$, then the reduced states are
\begin{equation} \label{marg1}
\hat{\sigma}_1^{(\mathrm{sym})} = \hat{\sigma}_2^{(\mathrm{sym})} = \frac{\ket{L} \bra{L} + \ket{R} \bra{R}}{2}.
\end{equation}
Note that the operational definition \eqref{margsym} only necessitates that $\hat{\sigma}_1 + \hat{\sigma}_2 = \ket{L} \bra{L} + \ket{R} \bra{R}$, which leaves some freedom in choosing the reduced states, e.g., as $\hat{\sigma}_1 \neq \hat{\sigma}_2$.
Motivated by the intuitive notion of particles, Ref.~\cite{Dieks2020} used this freedom to devise an interpretation of the quantum mechanics of indistinguishable particles in which $\hat{\sigma}_1 = \ket{L} \bra{L}$ and $\hat{\sigma}_2 = \ket{R} \bra{R}$. This interpretation, however, breaks down for bosons whenever $\av{L | R} \neq 0$ (for fermions, the interpretation requires amendments but does not fall apart overall). Moreover, specific assignments for reduced states do not entail any observational consequences, since, due to their symmetry, measurements cannot verify whether $\hat{\sigma}_1^{(\mathrm{sym})} = \hat{\sigma}_2^{(\mathrm{sym})}$ as in Eq.~\eqref{marg1} or $\hat{\sigma}_1 = \ket{L} \bra{L}$ and $\hat{\sigma}_2 = \ket{R} \bra{R}$.

When the joint state is not permutation-symmetric, there is no reason to assume that $\hat{\sigma}_1 = \hat{\sigma}_2$.
For example, when the joint state is $\ket{L} \otimes \ket{R}$, Eq.~\eqref{margsym} implies that $\hat{\sigma}_1 + \hat{\sigma}_2 = \ket{L} \bra{L} + \ket{R} \bra{R}$ for arbitrary states $\ket{L}$ and $\ket{R}$, not necessarily orthogonal. Now it is most sensible to assign $\hat{\sigma}_1 = \ket{L} \bra{L}$ and $\hat{\sigma}_2 = \ket{R} \bra{R}$ which is a prescription that we follow. Operationally, this prescription is justified by the fact that product states such as $\ket{L} \otimes \ket{R}$ produce independent probabilities with respect to measurements of the form $\hat{O} \otimes \hat{O}$; such an example is provided by the position measurement in Eq.~\eqref{bebe1}.

\section{Eigenvalue degeneracy}
\label{app:degeneracy}

To find the eigenstates of operator $\hat{t}$, we look for the solutions of the equation 
$\hat{t} \ket{\phi} = t_c \ket{\phi}$, where $t_c$ denotes the eigenvalue of $\hat{t}$. 
In the $u$-representation, according to Eq.~\eqref{eq:tquant}, this equation takes the form
\bea \label{plyum}
\phi'(u) = \frac{\phi(u)}{2 u} + \frac{i t_c u}{2 m} \phi(u) .
\eea
We seek the solutions among the continuous, piecewise differentiable functions $\phi(u) = \av{u | \phi}$. 
Solving Eq.~\eqref{plyum} for $u < 0$ and $u > 0$ gives
\bea \label{puk}
\phi(u) = \bigg\{\begin{array}{ll}
C_- \sqrt{-u} \, e^{i t_c u^2 / (4 m)}, & u < 0 \\
C_+ \sqrt{u} \, e^{i t_c u^2 / (4 m)}, & u > 0
\end{array},
\eea
where $C_-$ and $C_+$ are some constants [cf. Eq.~\eqref{rok}]. 
The solutions of Eq.~\eqref{plyum} are unique both for $u < 0$ and $u > 0$. 
Therefore, if Eq.~\eqref{plyum}, along with the continuity and piecewise differentiability of $\phi(u)$, 
provides a way to connect $C_-$ and $C_+$, the eigenstate will be unique. 
Otherwise, we shall extend the solutions \eqref{puk} to (resp.) $u < 0$ and $u > 0$ as in Eq.~\eqref{rok}, and the eigenstate will be double degenerate.

The only point at which the two solutions meet is $u = 0$. We see that $\phi(0-) = \phi(0+) = 0$, so the continuity condition does not provide sufficient information on $C_-$ and $C_+$, since $\phi(0) = 0$ for any $C_-$ and $C_+$. As for $\phi'(u)$, the eigenvalue equation \eqref{plyum} leads to
\begin{equation} \label{bum}
\phi'(0-) = \lim_{\epsilon_1 \to 0-} \frac{C_-}{\sqrt{\epsilon_1}}, \quad \phi'(0+) = \lim_{\epsilon_2 \to 0+} \frac{C_+}{\sqrt{\epsilon_2}} ,
\end{equation}
which again does not connect $C_-$ to $C_+$ in any conceivable way. Indeed, if we, e.g., impose $\phi'(0-) = \phi'(0+)$, we see that $C_- = C_+$ when $\epsilon_1 = \epsilon_2 \to 0$, $C_- = 2 \, C_+$ when $\epsilon_1 = 4 \epsilon_2 \to 0$, and so on, which means that $C_-$ and $C_+$ are independent. In a sense, the singularity of $\hat{t}$ at $u = 0$ is too strong to allow to uniquely connect the eigensolutions from both sides. The independence of $C_-$ and $C_+$ means double degeneracy, and the functions in Eq.~\eqref{rok} constitute an orthogonal basis in the eigensubspace of the eigenvalue $t_c$.

\section{Nonexistence of collision time observable}
\label{app:time}

The physical reason for non-existence of a self-adjoint time operator is that the possible energetic states of a system (the spectrum of the Hamiltonian) has to be limited from below: if there were a self-adjoint time operator, it could be used to translate energy (as momentum is used to translate coordinate), yielding physical states with arbitrarily low energies \cite{Muga1998, Leon2017}.

To illustrate this statement, let us assume that there exists a collision time observable $\hat{t}_c$, and consider its expectation value obtained from measurements on the two-particle system in some initial state $\ket{\Psi}$: $\bra{\Psi} \hat{t}_c \ket{\Psi}$. If we start observing the system not at time $t = 0$, but at a later time $t > 0$, the collision time should be reduced by $t$. We therefore impose the time-translation rule $\bra{\Psi (t)} \hat{t}_c \ket{\Psi (t)} = \bra{\Psi} \hat{t}_c \ket{\Psi} - t$, which is equivalent to 
\begin{equation} \label{pauli1}
e^{i t (H \otimes \id + \id \otimes H)} \hat{t}_c e^{- i t (H \otimes \id + \id \otimes H)} = \hat{t}_c - t .
\end{equation}
Since this expression should hold for any $t$, it is equivalent to 
\begin{equation} \label{pauli2}
[H \otimes \id + \id \otimes H, \hat{t}_c] = i,
\end{equation}
which, as mentioned above, is incompatible with $H \otimes \id + \id \otimes H$ having a spectrum bounded from below. Therefore, the collision time operator $\hat{t}_c$ cannot be a self-adjoint operator that satisfies the Born's rule.

\section{Features of POVM}
\label{app:povm}

POVMs represent (generalized) quantum measurements and determine the probabilities of measurement outcomes through the Born rule \cite{Peres}. This observation is used in the literature \cite{Mugaetal} to postulate that the outcome probabilities of measuring $\hat{t}$ defined in Eq.~\eqref{eq:tquant} are determined by the POVM comprised of the operators
$\ket{t_c ,s} \bra{t_c, s}$ ($s = \pm$), which are Hermitian, positive-semidefinite, and sum up to the identity: $\sum_s \int \d t_c \ket{t_c ,s} \bra{t_c, s} = \id^{(u)}$. Hence, in view of Eq.~\eqref{eq:tcfromt}, the POVM describing the measurement of $\hat{t}_c$ will be comprised of operators
\begin{equation} \label{buran2}
\ket{t_c ,s} \bra{t_c, s} \circ \id^{(v)}, \qquad s = \pm.
\end{equation}

A POVM, while being able to determine the probabilities of measurement outcomes, 
does not give a unique prescription (operator) to identify the post-measurement states. More specifically, if $\id \geq \hat{M}_x \geq 0$ is an element of a POVM, then the most general reduction rule for it can be written as
\bea \label{eq:redu}
\hat{\sigma} \to \frac{\hat{A}_x \hat{\sigma} \hat{A}_x^\dagger}{\tr(\hat{\sigma} M_x)},
\eea
where $\hat{\sigma}$ is a density operator and the otherwise arbitrary operator $\hat{A}_x$ satisfies $\hat{A}_x^\dagger \hat{A}_x = \hat{M}_x$ \cite{mikeike}. In our case [Eq.~\eqref{buran2}], the most general such decomposition is
\begin{eqnarray} \label{zair}
\ket{t_c ,s} \bra{t_c, s} \circ \id^{(v)} = \big[ \ket{t_c, s} \bra{\phi_{s, t_c}} \circ \id^{(v)} \big] \hat{V}_{s, t_c}^\dagger \cdot \hat{V}_{s, t_c} \big[ \ket{\phi_{s, t_c}} \bra{t_c, s} \circ \id^{(v)} \big],
\end{eqnarray}
where $\ket{\phi_{s, t_c}}$ are arbitrary normalized states, $\braket{\phi_{s, t_c} | \phi_{s, t_c}} = 1$, and $\hat{V}_{s, t_c}$ are arbitrary unitary operators ($\hat{V}_{s, t_c}^\dagger \hat{V}_{s, t_c} = \id^{(u)} \circ \id^{(v)}$).
Note that, in view of their normalization, states $\ket{\phi_{s, t_c}}$ cannot coincide with $\ket{t_c, s}$ because of Eq.~\eqref{nigeria}.

Thus, the general state-reduction rule described by Eq.~\eqref{eq:redu} for POVM elements given by Eq.~\eqref{buran2} is
\begin{equation} \label{kongo0}
\hat{\sigma} \to \frac{\hat{V}_{s, t_c} \big[ \ket{\phi_{s, t_c}} \bra{t_c, s} \circ \id^{(v)} \big] \hat{\sigma} \big[ \ket{t_c, s} \bra{\phi_{s, t_c}} \circ \id^{(v)} \big] \hat{V}_{s, t_c}^\dagger} 
{\tr \left(\hat{\sigma} \big[ \ket{t_c ,s} \bra{t_c, s} \circ \id^{(v)} \big] \right)}.
\end{equation}
Introducing the $v$-space operator $\braket{t_c, s | \hat{\sigma} | t_c, s} = \sum_{n, n'} [\bra{t_c, s} \circ \bra{n}] \, \hat{\sigma} \, [\ket{t_c, s} \circ \ket{n'}] \cdot \ket{n}\bra{n'}$, where $\{ \ket{n} \}_n$ is an arbitrary basis in the $v$-space, we can rewrite Eq.~\eqref{kongo0} as
\begin{equation} \label{kongo}
\hat{\sigma} \to \frac{\hat{V}_{s, t_c} \big[ \ket{\phi_{s, t_c}} \bra{\phi_{s, t_c}} \circ \braket{t_c, s | \hat{\sigma} | t_c, s} \big] \hat{V}_{s, t_c}^\dagger} 
{\tr \big(\braket{t_c, s | \hat{\sigma} | t_c, s}\big)}.
\end{equation}
When initial state $\hat{\sigma}$ is a pure state, Eq.~\eqref{kongo} simply reduces to
\bea \label{kongo1}
\hat{\sigma} \to \ket{\xi_{s, t_c}} \bra{\xi_{s, t_c}},
\eea
where
\bea
\ket{\xi_{s, t_c}} = \hat{V}_{s, t_c} \big[ \ket{\phi_{t_c, s}} \circ \ket{\varsigma} \big],
\eea
with
\bea
\ket{\varsigma} = \sum_n \frac{\big[ \bra{t_c, s} \circ \bra{n} \big] \ket{\xi_{s, t_c}}}{\sqrt{\tr \big(\braket{t_c, s | \hat{\sigma} | t_c, s}\big)}} \, \ket{n};
\eea
the state $\ket{\varsigma}$ is normalized and does not depend on the choice of the basis $\{\ket{n} \}_n$.

Since $\hat{V}_{s, t_c}$ is arbitrary, $\ket{\xi_{s, t_c}}$ can be an arbitrary pure state that does not depend on $s$. In particular, it can be chosen to be $\ket{\Psi_{\mathrm{s}}}$ of Eq.~\eqref{eq:Psisym}, so the quantum jump in Eq.~\eqref{eq:qjump} can be made consistent with the post-measurement state change \eqref{kongo1} induced by the collision time operator $\hat{t}_c$.

Note that Eq.~\eqref{kongo} does not violate the repeatability
principle: if the same measurement is carried out
twice, then the second measurement should confirm the result of the
first one. This is because there exists an actual projective measurement in a larger Hilbert space the reduction of which to the Hilbert space of the two particles induces the POVM in Eq.~\eqref{buran2} and the transition in Eq.~\eqref{kongo} \cite{Peres}

\section{Detection of Gaussian wavepackets}
\label{app:Gauss}

We assume the Gaussian wavepackets of Eqs.~\eqref{eq:psipt}, \eqref{eq:chip}, \eqref{gustaf}, \eqref{eq:psixt} with $d \equiv c_R - c_L > 0$, $a \equiv a_R = a_L$, and $b \equiv b_L = -b_R$. The relevant quantities in Eq.~\eqref{eq:Trsigt} are then given by the following explicit expressions:
\begin{eqnarray} \nonumber
\braket{R(t) | \hat{P} | L(t)} &=& \int_{(c_R+c_L)/2 - \ell}^{(c_R+c_L)/2 + \ell} \d x \, \psi_R^*(x,t)\psi_L(x,t)
\\ \nonumber 
&=& e^{-b^2-\frac{d^2}{4a^2}} \int_{- \frac{\ell}{a\sqrt{1+\tau^2}}}^{ \frac{\ell}{a\sqrt{1+\tau^2}}} \frac{\d x}{\sqrt{\pi}} \,
\exp \bigg[-\bigg( x - i \frac{\frac{\tau d}{2a}+b}{\sqrt{1+\tau^2}}\bigg)^2\bigg] 
\\ \nonumber  
&=& \frac{1}{2}\,e^{-b^2-\frac{d^2}{4a^2}}\left[ \erf \left( \frac{\ell-i(ab+\tau(d/2)\,)}{a\sqrt{1+\tau^2}} \right) - \erf \left(- \frac{\ell+i(ab+\tau (d/2)\,)}{a\sqrt{1+\tau^2}} \right)\right], \label{gott}
\\ \nonumber 
\braket{R(t) | \hat{P} | R(t)} &=& \frac{1}{2}\left[ \erf \left( \frac{\ell+ab\tau- (c/2) }{a\sqrt{1+\tau^2}} \right) - \erf \left( \frac{-\ell+ab\tau- (d/2) }{a\sqrt{1+\tau^2}} \right) \right],
\\ \nonumber 
\braket{L(t) | \hat{P} | L(t)} &=& \frac{1}{2}\left[ \erf \left( \frac{\ell-ab\tau+ (d/2) }{a\sqrt{1+\tau^2}} \right) - \erf \left( \frac{-\ell-ab\tau+ (d/2) }{a\sqrt{1+\tau^2}} \right) \right], \label{vandal}
\end{eqnarray}
where $\erf(x) \equiv \frac{2}{\sqrt{\pi}} \int_0^\infty \d s \, e^{-s^2}$ is the error function defined 
with convention $\erf(x) \to 1$ for $x \to \infty$, and $\tau = t / (m a^2)$ is the dimensionless time. For $\ell\to\infty$ we get $\braket{R | \hat{P} | L} \to e^{-b^2-\frac{d^2}{4a^2}}$ which is consistent with Eq.~\eqref{eq:ovrlpLR}.

\section{3D collisions}
\label{app:3D}

In the main text, we considered a 1D model, where the physical picture of inter-particle collisions is rather clear and can be analyzed in both semiclassical and the fully quantum pictures. In particular, we neglected inter-particle interactions having in mind that including a short-range inter-particle potential is straightforward. It amounts to redefining the collision operator from coinciding coordinates to the coordinate difference equal to the characteristic length $l$ of a short-range inter-particle potential. In 3D the situation is more complicated for at least two reasons: First the inter-particle interaction \textit{cannot} be neglected, since the point coordinates of (classical) particles generically never coincide in $3$D, i.e., the set of initial conditions for which this happens is of measure zero. This means that in 3D a finite characteristic range $l$ of the interaction potential should always be present. Second, the classical expression for the collision time presented below is conditional and complicated, and its quantization is as yet unclear. Hence, presently, we can estimate the collision time in $3$D only semiclassically. 

Considering two classical particles with a free dynamics up to the relative distance $l$, we have 
\begin{equation}
\vec{x}_k(t)=\vec{x}_k(0)+\frac{\vec{p}_k(0)t}{m}, \quad k=1,2,
\end{equation}
and the collision time is determined from 
\begin{equation} \label{co}
|\vec{x}_1(t)-\vec{x}_2(t)| = l,
\end{equation} 
which implies a spherically symmetric inter-particle potential with range $l>0$. 
Solutions of the quadratic equation \eqref{co} are analyzed assuming that $|\vec{x}_1(0)-\vec{x}_2(0)|>l$ at the initial time $t=0$. 
Out of the two solutions of Eq.~\eqref{co}, we should select the one where $t$ is positive and closest to zero. 
Denoting by $\vec{x}_{12}=\vec{x}_1(0)-\vec{x}_2(0)$ and $\vec{p}_{12}=\vec{p}_1(0)-\vec{p}_2(0)$, 
we obtain that the collision happens under the conditions
\begin{equation} 
\label{rashid10}
-\vec{x}_{12}\cdot \vec{p}_{12}>|\vec{p}_{12}|\sqrt{\vec{x}_{12}^2-l^2}>0,
\end{equation}
while the collision time is
\begin{equation} 
\label{rashid20}
t_\mr{c} = -\frac{1}{m \vec{p}_{12}^{\,2}} 
\left[\vec{x}_{12}\cdot \vec{p}_{12}+\sqrt{(\vec{x}_{12}\cdot \vec{p}_{12})^2 - \vec{p}_{12}^{\,2}(\vec{x}_{12}^{\,2}-l^2)} \right]. 
\end{equation}
Note that the first inequality in Eq.~\eqref{rashid10} ceases to hold for $l \to 0$, which confirms the necessity of a finite interaction range $l$.

Equations \eqref{rashid10} and \eqref{rashid20} suffice for making semiclassical estimates of the collision times using Wigner functions [cf. Eq.~\eqref{eq:rhocl}], but their quantization requires further research.

\bibliography{references}

\begin{thebibliography}{32}%
\makeatletter
\providecommand \@ifxundefined [1]{%
 \@ifx{#1\undefined}
}%
\providecommand \@ifnum [1]{%
 \ifnum #1\expandafter \@firstoftwo
 \else \expandafter \@secondoftwo
 \fi
}%
\providecommand \@ifx [1]{%
 \ifx #1\expandafter \@firstoftwo
 \else \expandafter \@secondoftwo
 \fi
}%
\providecommand \natexlab [1]{#1}%
\providecommand \enquote  [1]{``#1''}%
\providecommand \bibnamefont  [1]{#1}%
\providecommand \bibfnamefont [1]{#1}%
\providecommand \citenamefont [1]{#1}%
\providecommand \href@noop [0]{\@secondoftwo}%
\providecommand \href [0]{\begingroup \@sanitize@url \@href}%
\providecommand \@href[1]{\@@startlink{#1}\@@href}%
\providecommand \@@href[1]{\endgroup#1\@@endlink}%
\providecommand \@sanitize@url [0]{\catcode `\\12\catcode `\$12\catcode
  `\&12\catcode `\#12\catcode `\^12\catcode `\_12\catcode `\%12\relax}%
\providecommand \@@startlink[1]{}%
\providecommand \@@endlink[0]{}%
\providecommand \url  [0]{\begingroup\@sanitize@url \@url }%
\providecommand \@url [1]{\endgroup\@href {#1}{\urlprefix }}%
\providecommand \urlprefix  [0]{URL }%
\providecommand \Eprint [0]{\href }%
\providecommand \doibase [0]{https://doi.org/}%
\providecommand \selectlanguage [0]{\@gobble}%
\providecommand \bibinfo  [0]{\@secondoftwo}%
\providecommand \bibfield  [0]{\@secondoftwo}%
\providecommand \translation [1]{[#1]}%
\providecommand \BibitemOpen [0]{}%
\providecommand \bibitemStop [0]{}%
\providecommand \bibitemNoStop [0]{.\EOS\space}%
\providecommand \EOS [0]{\spacefactor3000\relax}%
\providecommand \BibitemShut  [1]{\csname bibitem#1\endcsname}%
\let\auto@bib@innerbib\@empty
\bibitem [{\citenamefont {Girardeau}(1965)}]{Girardeau1965}%
  \BibitemOpen
  \bibfield  {author} {\bibinfo {author} {\bibfnamefont {M.~D.}\ \bibnamefont
  {Girardeau}},\ }\bibfield  {title} {\bibinfo {title} {Permutation symmetry of
  many-particle wave functions},\ }\href
  {https://doi.org/10.1103/PhysRev.139.B500} {\bibfield  {journal} {\bibinfo
  {journal} {Phys. Rev.}\ }\textbf {\bibinfo {volume} {139}},\ \bibinfo {pages}
  {B500} (\bibinfo {year} {1965})}\BibitemShut {NoStop}%
\bibitem [{\citenamefont {Flicker}\ and\ \citenamefont
  {Leff}(1967)}]{Flicker1967}%
  \BibitemOpen
  \bibfield  {author} {\bibinfo {author} {\bibfnamefont {M.}~\bibnamefont
  {Flicker}}\ and\ \bibinfo {author} {\bibfnamefont {H.~S.}\ \bibnamefont
  {Leff}},\ }\bibfield  {title} {\bibinfo {title} {Symmetrization postulate of
  quantum mechanics},\ }\href {https://doi.org/10.1103/PhysRev.163.1353}
  {\bibfield  {journal} {\bibinfo  {journal} {Phys. Rev.}\ }\textbf {\bibinfo
  {volume} {163}},\ \bibinfo {pages} {1353} (\bibinfo {year}
  {1967})}\BibitemShut {NoStop}%
\bibitem [{\citenamefont {Salzman}(1970)}]{Salzman1970}%
  \BibitemOpen
  \bibfield  {author} {\bibinfo {author} {\bibfnamefont {W.~R.}\ \bibnamefont
  {Salzman}},\ }\bibfield  {title} {\bibinfo {title} {Exchange symmetry of
  many-particle state functions},\ }\href
  {https://doi.org/10.1103/PhysRevA.2.1664} {\bibfield  {journal} {\bibinfo
  {journal} {Phys. Rev. A}\ }\textbf {\bibinfo {volume} {2}},\ \bibinfo {pages}
  {1664} (\bibinfo {year} {1970})}\BibitemShut {NoStop}%
\bibitem [{\citenamefont {Peres}(1993)}]{Peres}%
  \BibitemOpen
  \bibfield  {author} {\bibinfo {author} {\bibfnamefont {A.}~\bibnamefont
  {Peres}},\ }\href@noop {} {\emph {\bibinfo {title} {Quantum Theory:
  {C}oncepts and Methods}}}\ (\bibinfo  {publisher} {Kluwer, Dordrecht},\
  \bibinfo {year} {1993})\BibitemShut {NoStop}%
\bibitem [{\citenamefont {Landau}\ and\ \citenamefont
  {Lifshitz}(1981)}]{Landau}%
  \BibitemOpen
  \bibfield  {author} {\bibinfo {author} {\bibfnamefont {L.~D.}\ \bibnamefont
  {Landau}}\ and\ \bibinfo {author} {\bibfnamefont {E.~M.}\ \bibnamefont
  {Lifshitz}},\ }\href@noop {} {\emph {\bibinfo {title} {Quantum Mechanics:
  {N}on-Relativistic Theory}}}\ (\bibinfo  {publisher} {Pergamon, London},\
  \bibinfo {year} {1981})\BibitemShut {NoStop}%
\bibitem [{\citenamefont {Mirman}(1973)}]{Mirman1973}%
  \BibitemOpen
  \bibfield  {author} {\bibinfo {author} {\bibfnamefont {R.}~\bibnamefont
  {Mirman}},\ }\bibfield  {title} {\bibinfo {title} {Experimental meaning of
  the concept of identical particles},\ }\href
  {https://doi.org/10.1007/BF02832643} {\bibfield  {journal} {\bibinfo
  {journal} {Nuovo Cimento B}\ }\textbf {\bibinfo {volume} {18}},\ \bibinfo
  {pages} {110} (\bibinfo {year} {1973})}\BibitemShut {NoStop}%
\bibitem [{\citenamefont {Gelfer}\ \emph {et~al.}(1975)\citenamefont {Gelfer},
  \citenamefont {Lyuboshitz},\ and\ \citenamefont {Podgoretskii}}]{Gelfer1975}%
  \BibitemOpen
  \bibfield  {author} {\bibinfo {author} {\bibfnamefont {Y.~M.}\ \bibnamefont
  {Gelfer}}, \bibinfo {author} {\bibfnamefont {L.~M.}\ \bibnamefont
  {Lyuboshitz}},\ and\ \bibinfo {author} {\bibfnamefont {I.~M.}\ \bibnamefont
  {Podgoretskii}},\ }\href@noop {} {\emph {\bibinfo {title} {Gibbs paradox and
  indistinguishability of particles in quantum mechanics}}}\ (\bibinfo
  {publisher} {Nauka, Moscow},\ \bibinfo {year} {1975})\ \bibinfo {note} {in
  Russian}\BibitemShut {NoStop}%
\bibitem [{\citenamefont {de~Muynck}\ and\ \citenamefont {van
  Liempd}(1986)}]{Muynck1986}%
  \BibitemOpen
  \bibfield  {author} {\bibinfo {author} {\bibfnamefont {W.~M.}\ \bibnamefont
  {de~Muynck}}\ and\ \bibinfo {author} {\bibfnamefont {G.~P.}\ \bibnamefont
  {van Liempd}},\ }\bibfield  {title} {\bibinfo {title} {On the relation
  between indistinguishability of identical particles and (anti)symmetry of the
  wave function in quantum mechanics},\ }\href
  {https://doi.org/10.1007/BF00485944} {\bibfield  {journal} {\bibinfo
  {journal} {Synthese}\ }\textbf {\bibinfo {volume} {67}},\ \bibinfo {pages}
  {477} (\bibinfo {year} {1986})}\BibitemShut {NoStop}%
\bibitem [{\citenamefont {Dieks}(1990)}]{Dieks1990}%
  \BibitemOpen
  \bibfield  {author} {\bibinfo {author} {\bibfnamefont {D.}~\bibnamefont
  {Dieks}},\ }\bibfield  {title} {\bibinfo {title} {Quantum statistics,
  identical particles and correlations},\ }\href
  {https://doi.org/10.1007/BF00413672} {\bibfield  {journal} {\bibinfo
  {journal} {Synthese}\ }\textbf {\bibinfo {volume} {82}},\ \bibinfo {pages}
  {127} (\bibinfo {year} {1990})}\BibitemShut {NoStop}%
\bibitem [{\citenamefont {Yunger~Halpern}\ and\ \citenamefont
  {Crosson}(2019)}]{YungerHalpern2019}%
  \BibitemOpen
  \bibfield  {author} {\bibinfo {author} {\bibfnamefont {N.}~\bibnamefont
  {Yunger~Halpern}}\ and\ \bibinfo {author} {\bibfnamefont {E.}~\bibnamefont
  {Crosson}},\ }\bibfield  {title} {\bibinfo {title} {Quantum information in
  the {P}osner model of quantum cognition},\ }\href
  {https://doi.org/10.1016/j.aop.2018.11.016} {\bibfield  {journal} {\bibinfo
  {journal} {Ann. Phys.}\ }\textbf {\bibinfo {volume} {407}},\ \bibinfo {pages}
  {92} (\bibinfo {year} {2019})}\BibitemShut {NoStop}%
\bibitem [{\citenamefont {Green}(1953)}]{Green1953}%
  \BibitemOpen
  \bibfield  {author} {\bibinfo {author} {\bibfnamefont {H.~S.}\ \bibnamefont
  {Green}},\ }\bibfield  {title} {\bibinfo {title} {A generalized method of
  field quantization},\ }\href {https://doi.org/10.1103/PhysRev.90.270}
  {\bibfield  {journal} {\bibinfo  {journal} {Phys. Rev.}\ }\textbf {\bibinfo
  {volume} {90}},\ \bibinfo {pages} {270} (\bibinfo {year} {1953})}\BibitemShut
  {NoStop}%
\bibitem [{\citenamefont {Ignatiev}\ and\ \citenamefont
  {Kuzmin}(1987)}]{Ignatiev1987}%
  \BibitemOpen
  \bibfield  {author} {\bibinfo {author} {\bibfnamefont {A.~Y.}\ \bibnamefont
  {Ignatiev}}\ and\ \bibinfo {author} {\bibfnamefont {V.~A.}\ \bibnamefont
  {Kuzmin}},\ }\bibfield  {title} {\bibinfo {title} {Is small violation of the
  {P}auli principle possible?},\ }\href@noop {} {\bibfield  {journal} {\bibinfo
   {journal} {Yad. Fiz.}\ }\textbf {\bibinfo {volume} {46}},\ \bibinfo {pages}
  {786} (\bibinfo {year} {1987})},\ \bibinfo {note} {available at
  \href{https://inis.iaea.org/collection/NCLCollectionStore/_Public/19/037/19037178.pdf}{URL}}\BibitemShut
  {NoStop}%
\bibitem [{\citenamefont {Messiah}(1962)}]{Messiah}%
  \BibitemOpen
  \bibfield  {author} {\bibinfo {author} {\bibfnamefont {A.}~\bibnamefont
  {Messiah}},\ }\href@noop {} {\emph {\bibinfo {title} {Quantum Mechanics}}},\
  Vol.~\bibinfo {volume} {2}\ (\bibinfo  {publisher} {North Holland,
  Amsterdam},\ \bibinfo {year} {1962})\BibitemShut {NoStop}%
\bibitem [{\citenamefont {Dalibard}\ \emph {et~al.}(1992)\citenamefont
  {Dalibard}, \citenamefont {Castin},\ and\ \citenamefont
  {M\o{}lmer}}]{Dalibard1992}%
  \BibitemOpen
  \bibfield  {author} {\bibinfo {author} {\bibfnamefont {J.}~\bibnamefont
  {Dalibard}}, \bibinfo {author} {\bibfnamefont {Y.}~\bibnamefont {Castin}},\
  and\ \bibinfo {author} {\bibfnamefont {K.}~\bibnamefont {M\o{}lmer}},\
  }\bibfield  {title} {\bibinfo {title} {Wave-function approach to dissipative
  processes in quantum optics},\ }\href
  {https://doi.org/10.1103/PhysRevLett.68.580} {\bibfield  {journal} {\bibinfo
  {journal} {Phys. Rev. Lett.}\ }\textbf {\bibinfo {volume} {68}},\ \bibinfo
  {pages} {580} (\bibinfo {year} {1992})}\BibitemShut {NoStop}%
\bibitem [{\citenamefont {Dum}\ \emph {et~al.}(1992)\citenamefont {Dum},
  \citenamefont {Zoller},\ and\ \citenamefont {Ritsch}}]{Dum1992}%
  \BibitemOpen
  \bibfield  {author} {\bibinfo {author} {\bibfnamefont {R.}~\bibnamefont
  {Dum}}, \bibinfo {author} {\bibfnamefont {P.}~\bibnamefont {Zoller}},\ and\
  \bibinfo {author} {\bibfnamefont {H.}~\bibnamefont {Ritsch}},\ }\bibfield
  {title} {\bibinfo {title} {Monte {C}arlo simulation of the atomic master
  equation for spontaneous emission},\ }\href
  {https://doi.org/10.1103/PhysRevA.45.4879} {\bibfield  {journal} {\bibinfo
  {journal} {Phys. Rev. A}\ }\textbf {\bibinfo {volume} {45}},\ \bibinfo
  {pages} {4879} (\bibinfo {year} {1992})}\BibitemShut {NoStop}%
\bibitem [{\citenamefont {Gardiner}\ \emph {et~al.}(1992)\citenamefont
  {Gardiner}, \citenamefont {Parkins},\ and\ \citenamefont
  {Zoller}}]{Gardiner1992}%
  \BibitemOpen
  \bibfield  {author} {\bibinfo {author} {\bibfnamefont {C.~W.}\ \bibnamefont
  {Gardiner}}, \bibinfo {author} {\bibfnamefont {A.~S.}\ \bibnamefont
  {Parkins}},\ and\ \bibinfo {author} {\bibfnamefont {P.}~\bibnamefont
  {Zoller}},\ }\bibfield  {title} {\bibinfo {title} {Wave-function quantum
  stochastic differential equations and quantum-jump simulation methods},\
  }\href {https://doi.org/10.1103/PhysRevA.46.4363} {\bibfield  {journal}
  {\bibinfo  {journal} {Phys. Rev. A}\ }\textbf {\bibinfo {volume} {46}},\
  \bibinfo {pages} {4363} (\bibinfo {year} {1992})}\BibitemShut {NoStop}%
\bibitem [{\citenamefont {Plenio}\ and\ \citenamefont
  {Knight}(1998)}]{Plenio1998}%
  \BibitemOpen
  \bibfield  {author} {\bibinfo {author} {\bibfnamefont {M.~B.}\ \bibnamefont
  {Plenio}}\ and\ \bibinfo {author} {\bibfnamefont {P.~L.}\ \bibnamefont
  {Knight}},\ }\bibfield  {title} {\bibinfo {title} {The quantum-jump approach
  to dissipative dynamics in quantum optics},\ }\href
  {https://doi.org/10.1103/RevModPhys.70.101} {\bibfield  {journal} {\bibinfo
  {journal} {Rev. Mod. Phys.}\ }\textbf {\bibinfo {volume} {70}},\ \bibinfo
  {pages} {101} (\bibinfo {year} {1998})}\BibitemShut {NoStop}%
\bibitem [{\citenamefont {Lambropoulos}\ and\ \citenamefont
  {Petrosyan}(2007)}]{Lambropoulos2007}%
  \BibitemOpen
  \bibfield  {author} {\bibinfo {author} {\bibfnamefont {P.}~\bibnamefont
  {Lambropoulos}}\ and\ \bibinfo {author} {\bibfnamefont {D.}~\bibnamefont
  {Petrosyan}},\ }\href@noop {} {\emph {\bibinfo {title} {Fundamentals of
  quantum optics and quantum information}}}\ (\bibinfo  {publisher} {Springer,
  Berlin},\ \bibinfo {year} {2007})\BibitemShut {NoStop}%
\bibitem [{\citenamefont {Hudson}(1974)}]{Hudson1974}%
  \BibitemOpen
  \bibfield  {author} {\bibinfo {author} {\bibfnamefont {R.~L.}\ \bibnamefont
  {Hudson}},\ }\bibfield  {title} {\bibinfo {title} {When is the {W}igner
  quasi-probability density non-negative?},\ }\href
  {https://doi.org/10.1016/0034-4877(74)90007-X} {\bibfield  {journal}
  {\bibinfo  {journal} {Rep. Math. Phys.}\ }\textbf {\bibinfo {volume} {6}},\
  \bibinfo {pages} {249} (\bibinfo {year} {1974})}\BibitemShut {NoStop}%
\bibitem [{\citenamefont {Aharonov}\ and\ \citenamefont
  {Bohm}(1961)}]{Aharonov1961}%
  \BibitemOpen
  \bibfield  {author} {\bibinfo {author} {\bibfnamefont {Y.}~\bibnamefont
  {Aharonov}}\ and\ \bibinfo {author} {\bibfnamefont {D.}~\bibnamefont
  {Bohm}},\ }\bibfield  {title} {\bibinfo {title} {Time in the quantum theory
  and the uncertainty relation for time and energy},\ }\href
  {https://doi.org/10.1103/PhysRev.122.1649} {\bibfield  {journal} {\bibinfo
  {journal} {Phys. Rev.}\ }\textbf {\bibinfo {volume} {122}},\ \bibinfo {pages}
  {1649} (\bibinfo {year} {1961})}\BibitemShut {NoStop}%
\bibitem [{\citenamefont {Grot}\ \emph {et~al.}(1996)\citenamefont {Grot},
  \citenamefont {Rovelli},\ and\ \citenamefont {Tate}}]{Grot1996}%
  \BibitemOpen
  \bibfield  {author} {\bibinfo {author} {\bibfnamefont {N.}~\bibnamefont
  {Grot}}, \bibinfo {author} {\bibfnamefont {C.}~\bibnamefont {Rovelli}},\ and\
  \bibinfo {author} {\bibfnamefont {R.~S.}\ \bibnamefont {Tate}},\ }\bibfield
  {title} {\bibinfo {title} {Time of arrival in quantum mechanics},\ }\href
  {https://doi.org/10.1103/PhysRevA.54.4676} {\bibfield  {journal} {\bibinfo
  {journal} {Phys. Rev. A}\ }\textbf {\bibinfo {volume} {54}},\ \bibinfo
  {pages} {4676} (\bibinfo {year} {1996})}\BibitemShut {NoStop}%
\bibitem [{\citenamefont {Muga}\ \emph {et~al.}(1998)\citenamefont {Muga},
  \citenamefont {Leavens},\ and\ \citenamefont {Palao}}]{Muga1998}%
  \BibitemOpen
  \bibfield  {author} {\bibinfo {author} {\bibfnamefont {J.~G.}\ \bibnamefont
  {Muga}}, \bibinfo {author} {\bibfnamefont {C.~R.}\ \bibnamefont {Leavens}},\
  and\ \bibinfo {author} {\bibfnamefont {J.~P.}\ \bibnamefont {Palao}},\
  }\bibfield  {title} {\bibinfo {title} {Space-time properties of free motion
  time-of-arrival eigenstates},\ }\href
  {https://doi.org/10.1103/PhysRevA.58.4336} {\bibfield  {journal} {\bibinfo
  {journal} {Phys. Rev. A}\ }\textbf {\bibinfo {volume} {58}},\ \bibinfo
  {pages} {4336} (\bibinfo {year} {1998})}\BibitemShut {NoStop}%
\bibitem [{\citenamefont {Fr\"{o}wis}\ \emph {et~al.}(2018)\citenamefont
  {Fr\"{o}wis}, \citenamefont {Sekatski}, \citenamefont {D\"{u}r},
  \citenamefont {Gisin},\ and\ \citenamefont {Sangouard}}]{Frowis2018}%
  \BibitemOpen
  \bibfield  {author} {\bibinfo {author} {\bibfnamefont {F.}~\bibnamefont
  {Fr\"{o}wis}}, \bibinfo {author} {\bibfnamefont {P.}~\bibnamefont
  {Sekatski}}, \bibinfo {author} {\bibfnamefont {W.}~\bibnamefont {D\"{u}r}},
  \bibinfo {author} {\bibfnamefont {N.}~\bibnamefont {Gisin}},\ and\ \bibinfo
  {author} {\bibfnamefont {N.}~\bibnamefont {Sangouard}},\ }\bibfield  {title}
  {\bibinfo {title} {Macroscopic quantum states: {M}easures, fragility, and
  implementations},\ }\href {https://doi.org/10.1103/RevModPhys.90.025004}
  {\bibfield  {journal} {\bibinfo  {journal} {Rev. Mod. Phys.}\ }\textbf
  {\bibinfo {volume} {90}},\ \bibinfo {pages} {025004} (\bibinfo {year}
  {2018})}\BibitemShut {NoStop}%
\bibitem [{\citenamefont {Muga}\ \emph {et~al.}(2002)\citenamefont {Muga},
  \citenamefont {Mayato},\ and\ \citenamefont {Egusquiza}}]{Mugaetal}%
  \BibitemOpen
  \bibinfo {editor} {\bibfnamefont {J.~G.}\ \bibnamefont {Muga}}, \bibinfo
  {editor} {\bibfnamefont {R.}~\bibnamefont {Mayato}},\ and\ \bibinfo {editor}
  {\bibfnamefont {I.~L.}\ \bibnamefont {Egusquiza}},\ eds.,\ \href@noop {}
  {\emph {\bibinfo {title} {Time in Quantum Mechanics}}},\ \bibinfo {series}
  {Lecture Notes in Physics Vol. 734}, Vol.~\bibinfo {volume} {1}\ (\bibinfo
  {publisher} {Springer, Heidelberg},\ \bibinfo {year} {2002})\BibitemShut
  {NoStop}%
\bibitem [{\citenamefont {Egusquiza}\ and\ \citenamefont
  {Muga}(1999)}]{Egusquiza1999}%
  \BibitemOpen
  \bibfield  {author} {\bibinfo {author} {\bibfnamefont {I.~L.}\ \bibnamefont
  {Egusquiza}}\ and\ \bibinfo {author} {\bibfnamefont {J.~G.}\ \bibnamefont
  {Muga}},\ }\bibfield  {title} {\bibinfo {title} {Free-motion time-of-arrival
  operator and probability distribution},\ }\href
  {https://doi.org/10.1103/PhysRevA.61.012104} {\bibfield  {journal} {\bibinfo
  {journal} {Phys. Rev. A}\ }\textbf {\bibinfo {volume} {61}},\ \bibinfo
  {pages} {012104} (\bibinfo {year} {1999})}\BibitemShut {NoStop}%
\bibitem [{\citenamefont {Paul}(1961)}]{Paul1961}%
  \BibitemOpen
  \bibfield  {author} {\bibinfo {author} {\bibfnamefont {H.}~\bibnamefont
  {Paul}},\ }\bibfield  {title} {\bibinfo {title} {\"{U}ber quantenmechanische
  {Z}eitoperatoren},\ }\href {https://doi.org/10.1002/andp.19624640505}
  {\bibfield  {journal} {\bibinfo  {journal} {Ann. Phys. (Berlin)}\ }\textbf
  {\bibinfo {volume} {9}},\ \bibinfo {pages} {252} (\bibinfo {year}
  {1961})}\BibitemShut {NoStop}%
\bibitem [{\citenamefont {Allcock}(1969)}]{Allcock1969}%
  \BibitemOpen
  \bibfield  {author} {\bibinfo {author} {\bibfnamefont {G.~R.}\ \bibnamefont
  {Allcock}},\ }\bibfield  {title} {\bibinfo {title} {The time of arrival in
  quantum mechanics {I}. formal considerations},\ }\href
  {https://doi.org/10.1016/0003-4916(69)90251-6} {\bibfield  {journal}
  {\bibinfo  {journal} {Ann. Phys.}\ }\textbf {\bibinfo {volume} {53}},\
  \bibinfo {pages} {253} (\bibinfo {year} {1969})}\BibitemShut {NoStop}%
\bibitem [{\citenamefont {Gieres}(2000)}]{Gieres2000}%
  \BibitemOpen
  \bibfield  {author} {\bibinfo {author} {\bibfnamefont {F.}~\bibnamefont
  {Gieres}},\ }\bibfield  {title} {\bibinfo {title} {Mathematical surprises and
  {D}irac's formalism in quantum mechanics},\ }\href
  {https://doi.org/10.1088/0034-4885/63/12/201} {\bibfield  {journal} {\bibinfo
   {journal} {Rep. Prog. Phys.}\ }\textbf {\bibinfo {volume} {63}},\ \bibinfo
  {pages} {1893} (\bibinfo {year} {2000})}\BibitemShut {NoStop}%
\bibitem [{\citenamefont {Leon}\ and\ \citenamefont
  {Maccone}(2017)}]{Leon2017}%
  \BibitemOpen
  \bibfield  {author} {\bibinfo {author} {\bibfnamefont {J.}~\bibnamefont
  {Leon}}\ and\ \bibinfo {author} {\bibfnamefont {L.}~\bibnamefont {Maccone}},\
  }\bibfield  {title} {\bibinfo {title} {The {P}auli objection},\ }\href
  {https://doi.org/10.1007/s10701-017-0115-2} {\bibfield  {journal} {\bibinfo
  {journal} {Found. Phys.}\ }\textbf {\bibinfo {volume} {47}},\ \bibinfo
  {pages} {1597} (\bibinfo {year} {2017})}\BibitemShut {NoStop}%
\bibitem [{\citenamefont {Giannitrapani}(1997)}]{Giannitrapani1997}%
  \BibitemOpen
  \bibfield  {author} {\bibinfo {author} {\bibfnamefont {R.}~\bibnamefont
  {Giannitrapani}},\ }\bibfield  {title} {\bibinfo {title}
  {Positive-operator-valued time observable in quantum mechanics},\ }\href
  {https://doi.org/10.1007/BF02435757} {\bibfield  {journal} {\bibinfo
  {journal} {Int. J. Theor. Phys.}\ }\textbf {\bibinfo {volume} {36}},\
  \bibinfo {pages} {1575} (\bibinfo {year} {1997})}\BibitemShut {NoStop}%
\bibitem [{\citenamefont {Dieks}\ and\ \citenamefont
  {Lubberdink}(2020)}]{Dieks2020}%
  \BibitemOpen
  \bibfield  {author} {\bibinfo {author} {\bibfnamefont {D.}~\bibnamefont
  {Dieks}}\ and\ \bibinfo {author} {\bibfnamefont {A.}~\bibnamefont
  {Lubberdink}},\ }\bibfield  {title} {\bibinfo {title} {Identical quantum
  particles as distinguishable objects},\ }\bibfield  {journal} {\bibinfo
  {journal} {J. Gen. Philos. Sci.}\ }\href
  {https://doi.org/10.1007/s10838-020-09510-w} {10.1007/s10838-020-09510-w}
  (\bibinfo {year} {2020})\BibitemShut {NoStop}%
\bibitem [{\citenamefont {Nielsen}\ and\ \citenamefont
  {Chuang}(2010)}]{mikeike}%
  \BibitemOpen
  \bibfield  {author} {\bibinfo {author} {\bibfnamefont {M.~A.}\ \bibnamefont
  {Nielsen}}\ and\ \bibinfo {author} {\bibfnamefont {I.~L.}\ \bibnamefont
  {Chuang}},\ }\href@noop {} {\emph {\bibinfo {title} {Quantum computation and
  quantum information}}}\ (\bibinfo  {publisher} {Cambridge University Press,
  Cambridge, England},\ \bibinfo {year} {2010})\BibitemShut {NoStop}%
\end{thebibliography}%

\end{document}